\documentclass{aa}  

\usepackage{graphicx}
\usepackage{txfonts}
\usepackage{multirow}
\usepackage[tbtags]{amsmath}
\begin{document}

   \title{MCMCI: A code to fully characterise an exoplanetary system}

   \subtitle{}

   \author{A. Bonfanti\inst{1}
          \and
          M. Gillon\inst{2}
          }

   \institute{Space Sciences, Technologies and Astrophysics Research (STAR) Institute, Université de Liège, 19C Allée du 6 Août, 4000 Liège, Belgium\\
              \email{a.bonfanti@uliege.be}
           \and
              Astrobiology Research Unit, Université de Liège, Allée du 6 Août 19, 4000 Liège, Belgium
             }

   \date{}

 
  \abstract
   {Useful information can be retrieved by analysing the transit light curve of a planet-hosting star or induced radial velocity oscillations. However, inferring the physical parameters of the planet, such as mass, size, and semi-major axis, requires preliminary knowledge of some parameters of the host star, especially its mass or radius, which are generally inferred through theoretical evolutionary models.}
   {We seek to present and test a whole algorithm devoted to the complete characterisation of an exoplanetary system thanks to the global analysis of photometric or radial velocity time series combined with observational stellar parameters derived either from spectroscopy or photometry.}
   {We developed an integrated tool called MCMCI. This tool combines the Markov chain Monte Carlo (MCMC) approach of analysing photometric or radial velocity time series with a proper interpolation within stellar evolutionary isochrones and tracks, known as isochrone placement, to be performed at each chain step, to retrieve stellar theoretical parameters such as age, mass, and radius.}
   {We tested the MCMCI on the HD 219134 multi-planetary system hosting two transiting rocky super Earths and on WASP-4, which hosts a bloated hot Jupiter. Even considering different input approaches, a final convergence was reached within the code, we found good agreement with the results already stated in the literature and we obtained more precise output parameters, especially concerning planetary masses.}
   {The MCMCI tool offers the opportunity to perform an integrated analysis of an exoplanetary system without splitting it into the preliminary stellar characterisation through theoretical models. Rather this approach favours a close interaction between light curve analysis and isochrones, so that the parameters recovered at each step of the MCMC enter as inputs for purposes of isochrone placement.}


   \maketitle
%

\section{Introduction}
A few decades have passed since the discovery of the first exoplanet orbiting the main-sequence (MS) star 51 Pegasi \citep{mayor95} and by now $\sim4000$ exoplanets have been confirmed, $\sim75\%$ of which are transiting planets\footnote{NASA Exoplanet Archive: \url{https://exoplanetarchive.ipac.caltech.edu/docs/counts_detail.html}}. 

In the initial years the most effective technique for discovering exoplanets was the radial velocity (RV) method \citep[see e.g.][]{wright18}, but during the last decade the transit method has become the most prominent, also thanks to past space missions, such as Kepler \citep{basri05,borucki10}, its extension K2 \citep{howell14}, and CoRoT \citep{rouan99,auvergne09}. In the meantime, other space telescopes, such as TESS \citep{sharma18} have just begun detecting exoplanetary transits; CHEOPS \citep{broeg13} has just been launched, whereas PLATO \citep{rauer14} is expected to be launched in 2026. 

As is well known, once the light curve (LC) of a planet-hosting star is available, the transit depth gives the squared ratio of planetary to stellar radii $\mathrm{d}F=\left(\frac{R_p}{R_{\star}} \right)^2$, while the orbital period $P$ and the transit shape parameters (impact parameter, depth, and duration) enable the retrieval of the stellar mean density $\rho_{\star}$. This is true for circular orbits under the assumption that exoplanetary mass $M_p$ is negligible if compared with stellar mass $M_{\star}$; otherwise RV measurements are needed to constrain  the orbital eccentricity $e$ and the planetary mass $M_p$ as well. A review of the transit technique is provided in \citet{winn10}.

From these considerations it is clear that establishing the exoplanetary parameters (in particular, retrieving $R_p$ in case we are just dealing with the transit method) requires not only the availability of transit photometry, but also knowledge of stellar parameters, either $M_{\star}$ or $R_{\star}$, as pointed out by \citet{seager03}. Thus, as a very basic example, LC analyses are usually split into three steps: first the LC-analyser tool recovers $\rho_{\star}$, after that $\rho_{\star}$ together with the stellar effective $T_{\mathrm{eff}}$ and metallicity [Fe/H] are employed to infer $M_{\star}$ and/or $R_{\star}$ from stellar evolutionary models, and finally the transit-analysis tool is launched again assuming also theoretical stellar parameters as inputs to determine $R_p$ \citep[see e.g.][]{gillon09b}.

In this paper we present our custom-developed MCMCI (Markov chain Monte Carlo + isochrones) \textsc{Fortran} code, which makes the transit analysis algorithm directly interact with stellar evolutionary models, so that starting from LCs and very basic stellar parameters, it is possible to characterise the whole exoplanetary system directly. Our MCMCI tool is very useful to carry out this kind of analysis and it will be very valuable, especially in coming years when lots of LCs and data from transits are expected to be available as a result of already active and forthcoming missions.

Our unified approach of simultaneously modelling the star and its planets is remarkable even if it is not unique, since it has already been followed, for example by \citet{hartman19}, who carry out an isochrone-based joint analysis using a differential evolution MCMC procedure \citep{terBraak06} and PARSEC stellar evolutionary models \citep{marigo17}. Also \citet{siverd12} simultaneously model the stellar host and its companion, but to break the $M_{\star}$-$R_{\star}$ degeneracy they simply consider the mass-radius relation by \citet{torres10}, as they use the first version of the \textsc{Idl} fitting package \texttt{EXOFAST} \citep{eastman13}. We notice that our version of MCMC without the support of stellar evolutionary models also allows us to select a modified version of the \citet{torres10} $M_{\star}$-$R_{\star}$ law from well-constrained detached binary systems \citep{gillon11}; thus we can use this tool during the LC or RV fitting.

Recently, a new version of the \texttt{EXOFAST} package called \texttt{EXOFASTv2} has been released \citep{eastman19}: it facilitates fitting the stellar properties along with the planetary fit with MIST evolutionary tracks \citep{dotter16} or Yonsei-Yale (YY) stellar evolutionary models \citep{yi01}, following the same philosophy as our MCMCI. However an optional spectral energy distribution (SED) fitting is also included, which our code does not perform. Moreover, besides jointly dealing with photometric and RV time series similar to our MCMCI, modelling stellar and planetary astrometric signals, modelling Doppler tomography \citep{collierCameron10}, and integrating a mass-radius relation for exoplanets are remarkable features of \texttt{EXOFASTv2} that are not implemented in MCMCI.

Other codes are available to address both the transit and RV fitting, such as \texttt{TLCM} \citep{smith17}, \texttt{PlanetPack3} \citep{baluev18}, \texttt{juliet} \citep{espinoza18}, \texttt{Pyaneti} \citep{barragan19}, \texttt{exoplanet}\footnote{\url{https://github.com/dfm/exoplanet}}, \texttt{allesfitter} \citep{guenther19}. The \texttt{TLCM} code is written in \textsc{Idl}, \texttt{PlanetPack3} is written in C++, while the other codes are written in \textsc{Python} (\texttt{Pyaneti} is also composed of \textsc{Fortran} routines, which are then wrapped to \textsc{Python}). A notable feature of \texttt{TLCM} is its possibility of simultaneously modelling RV of both components of a binary star, and it incorporates ellipsoidal modulation \citep{kopal59,morris85} and a beaming effects treatment \citep{rybicki79,loeb03} in the LCs. The \texttt{PlanetPack3} tool includes Gaussian processes \citep{rasmussen05} to model the noise in RV data, but does not allow us to fit multiple planets in the photometry. The \texttt{juliet}, \texttt{exoplanet}, and \texttt{allesfitter} codes account for Gaussian processes in both LC and RV time series using \texttt{celerite} \citep{foreman17}. Moreover, the core of \texttt{exoplanet} is the implementation of an Hamiltonian MCMC procedure \citep[see e.g.][]{neal12,betancourt17} that promises faster convergence time with respect to traditional MCMC implementation, while \texttt{allesfitter} also enables us to model star spots and stellar flares using \texttt{aflare} \citep{davenport14}.

Apart from \texttt{EXOFASTv2}, all the other listed codes do not include a simultaneous analysis of the stellar host through stellar evolutionary models along the LC and RV fitting. Our MCMCI, instead, allows this isochrone-based joint analysis and may also give different constraints for the stellar age (and thus the age of the entire exoplanetary system) using both model-dependent and empirical age indicators. We finally notice that our tool is unique in that it is fully implemented in \textsc{Fortran}.

In \S\ref{sec:code} an overview of the code is given, in \S\ref{sec:results} its application on two different exoplanetary systems is presented, and \S\ref{sec:conclusions} reports our conclusions.

\section{Code description}\label{sec:code}

The MCMCI is a \textsc{Fortran} program that was born from a proper merging of the custom MCMC code widely presented in \citet{gillon10,gillon12} and the isochrone placement algorithm, which is described in \citet{bonfanti15,bonfanti16}. In the following subsections we recall key aspects of the two codes according to their most recent updates and we explain how we merged these tools to facilitate their interaction.

\subsection{Markov chain Monte Carlo MCMC}\label{ssec:MCMC}
The MCMC simulation is a stochastic process \citep[see e.g.][]{holman06,ford05} that generates a sequence (chain) of data points (states) starting from an initial state that is then perturbed. The aim of this simulation is to sample the probability distribution function (PDF) of some parameters of interest assuming a given model; our version makes use of the Metropolis-Hastings algorithm \citep{hastings70} and the Gibbs sampler \citep[and references therein]{casella92}.

It is likely that the first states of a chain do not come from the limiting distribution, therefore a common practice in the MCMC approach is to discard these first data points: this process is called burn-in. In this way the effect of initial values on the posterior inference is minimised. Choice of the burn-in length depends upon the initial state and the speed of convergence to the limiting distribution. Establishing the proper burn-in length requires analysis of the output case by case, however experience suggests that setting the burn-in length to $20\%$ represents a conservative compromise. This is the value that has been chosen for all the analysis described in \S\ref{sec:results}. Anyway, our input form enables us to manually specify the length of the burn-in phase.

\subsubsection{Models and input parameters}
Our code may deal with any number of LC and RV time series for a complete joint analysis of the transit and dynamics of the exoplanetary system. If only LCs are available, only information recoverable from photometry is retrieved (e.g. the planetary radius $R_p$), and a determination of planetary mass $M_p$ is not possible unless the RV semi-amplitude $K$ is known from some RV studies. Vice versa, if only RV time series are available, the exoplanetary system are characterised from a dynamic point of view, that is $M_p$ is obtained, but not $R_p$. Our implementation of the code assumes use of the photometric model by \citet{mandel02} to reproduce the eclipse; and a classical Keplerian model for analysis of the RV signal, in addition to a Rossiter-McLaughlin effect model \citep{gimenez06} if RVs are obtained during transit. Our global model may deal with any number of planets, either transiting or not.

The eclipse model is multiplied by a trend model aimed to reproduce all the systematic effects having either astrophysical or instrumental origin, which are responsible for photometric variations beyond those caused by the transit itself. Specifically, through polynomial functions, this baseline is able to model, for example, the effect of inhomogeneous intra-pixel sensitivity of the detector \citep{knutson08,charbonneau08}; the so-called ramp effect, according to which the detector gain may increase asymptotically over time, depending on the illumination history of the pixels \citep{knutson08,charbonneau08}; the time-dependent photometric modulation of the stellar signal due to rotating spots; and the effect of the variations of the sky background, the airmass, or the width of the point spread function during observation.

The code is also able to face trends in the RV time series in terms of time, cross-correlation function width and bisector parameters \citep{baranne96}, and $\log{R'_{\mathrm{HK}}}$ \citep[a typical indicator of stellar magnetic activity; see e.g.][and references therein]{wright04}.

Barycentric Dynamic Time (TDB) is the default time standard that is used by the code in all the analyses; TDB is suitable for precision time monitoring. This time standard is preferable to the Coordinated Universal Time (UTC) because it avoids the drift due to leap seconds; in addition, it refers to the barycentre of the Solar System, thus it corrects for relativistic effect due to the gravitational potential of Earth. A discussion about time standards and timing precision is provided in for example \citet{eastman10}.

The following set of parameters may be randomly perturbed at each chain step (jump parameters or step parameters).
\begin{itemize}
\item The transit depth $\mathrm{d}F$.

\item  The occultation depth $\mathrm{d}F_{\mathrm{occ}}$.

\item  The impact parameter in the case of circular orbit $b'$.

\item  The eclipse duration $W$, that is the time between the first and last contact, as defined in \citet{winn10}.

\item  The eclipse timing $T_0$, that is the time of inferior conjunction, when the true anomaly at transit is $f_t=90^{\circ}-\omega$ (with $\omega$ argument of the periastron).

\item  The orbital period $P$.

\item  The quantities $\sqrt{e}\sin{\omega}$ and $\sqrt{e}\cos{\omega}$, where $e$ is the orbital eccentricity and $\omega$ is the argument of the periastron.
 
\item The parameter $K_2=K\sqrt{1-e^2}P^{1/3}$, where $K$ is the RV semi-amplitude.

\item The quantities $\sqrt{v\sin{i_{\star}}}\cos{\beta}$ and $\sqrt{v\sin{i_{\star}}}\sin{\beta}$, where $v\sin{i_{\star}}$ is the stellar rotational velocity along the line of sight and $\beta$ is the projected angle between the stellar spin axis and orbital axis.

\item The stellar metallicity [Fe/H]$_{\star}$.

\item The stellar effective temperature $T_{\mathrm{eff}}$ or colour index.

\item The stellar radius $R_{\star}$.

\item The stellar mass $M_{\star}$.

\item The limb-darkening (LD) coefficients.
\end{itemize}
 
If more than one planet has to be fitted, all the relevant parameters of each planet are listed in the input form.

The algorithm checks whether the drawn values of the jump parameters are physical. Therefore, for instance, negative values for $\mathrm{d}F$, $\mathrm{d}F_{\mathrm{occ}}$, $b'$, $W$, $T_0$, $P$, $K_2$, $T_{\mathrm{eff}}$, $R_{\star}$ or $M_{\star}$ are not allowed, any square root arguments must be positive, and the inequalities $|\sqrt{e}\cos{\omega}|<1$, $|\sqrt{e}\sin{\omega}|<1$ must hold since we are dealing with bound orbital systems. If any of the drawn value is not physical, the jump is not accepted, and a copy of the previous state is made.

\subsubsection{Comments on jump parameters}

Stepping in $\mathrm{d}F$, that is assuming a uniform prior on $\mathrm{d}F$, creates an implicit bias on $R_p=\sqrt{\mathrm{d}F}R_{\star}$ towards higher $R_p$ values. If the slope of the $R_p$ prior is flat enough over the range of interest of the posterior, this does not unphysically bias the posterior, however an alternative approach would be to replace the stepping parameter $\mathrm{d}F$ with $\frac{R_p}{R_{\star}}$, as proposed for example by \citet{eastman13}.

The impact parameter in case of circular orbit is given by $b'=a\cos{i_\mathrm{p}}/R_{\star}$, where $a$ is the semi-major axis of planetary orbit and $i_\mathrm{p}$ is the orbit inclination with respect to the plane of the sky. As pointed out by \citet{gillon09b}, assuming $b'$ rather than the actual impact parameter $b=b'\frac{1-e^2}{1+e\sin{\omega}}$ is preferable to minimise the correlation between jump parameters.
 
Our algorithm can also handle eccentric orbits by enabling the stepping in $\sqrt{e}\sin{\omega}$ and $\sqrt{e}\cos{\omega}$. This parametrisation ensures both orthogonality and uniform priors on $e$ and $\omega$, which has been pointed out for the first time by \citet{anderson11} and then broadly discussed by \citet{eastman13}.
In case RV data are available, the usual approach is to launch a first run with a fixed $e=0$ value, and then launch another run letting $e$ as a free jump parameter and check whether the BIC (eq. (\ref{eq:BIC}), see later) is in favour of an eccentric or circular solution.
 
 Instead, in case of a transit-only fit, treating $e$ is generally problematic. On the one hand, orbits of close-in exoplanets are likely circular. \citet{dawson18} report that for orbital periods $P<3$ days (i.e. orbital semi-major axes $a<0.04$ AU, if a Sun-like star is assumed as host), hot Jupiters are consistent with $e=0$ orbits, while instead in the $3<P<10$ range (in days, i.e. $0.04\textrm{AU}<a<0.09\textrm{AU}$) it is common to observe moderately elliptical orbits. This is consistent with the circularisation process induced by tidal evolution: the shortest the orbital period, the fastest the circularisation timescale. Moreover, the less massive the planet, the fastest the circularisation timescale. Considering an initial semi-major axis $a_0=0.04$ AU and an initial orbital eccentricity $e_0=0.1$, the orbital circularisation happens after $\sim150$ Myr, $\sim100$ Myr, and $\sim30$ Myr for a hot Jupiter, a hot Neptune, and a super-Earth, all orbiting a Sun-like star \citep{rodriguez10,rodriguez11}. As a consequence, less massive planets are expected to exhibit $e=0$ orbits at larger host star separation with respect to more massive planets \citep{pont11}. Summing up, it is reasonable to assume circular orbits for semi-major axes lower than a few hundredths of astronomical units.

 On the other hand, if we are dealing with outer exoplanets, setting $e=0$ may cause some biases on the derived parameters of the system. In this case, we could infer the mean stellar density as a result of evolutionary models (say $\rho_{\star\mathrm{,th}}$ its value) and use it together with the transit duration trying to constrain $e$. To do so, we should set $\rho_{\star\mathrm{,th}}$ as a prior in our MCMCI input form and let $P$, $\mathrm{d}F$, $W$, $b$, $e$, and $\omega$ as jump parameters. The transit-based $\frac{a}{R_{\star}}$ parameter, which is computed by the code through eq. (\ref{eq:aR}), is translated to $\rho_{\star}$ through Kepler's third law and then prior $\rho_{\star\mathrm{,th}}$ drives the $\rho_{\star}$ value. Thus, favoured values of the transit parameters are those which produce a $\rho_{\star}$ value more similar to $\rho_{\star\mathrm{,th}}$. The ($e$, $\omega$) degeneracy may not lead to a complete convergence of the transit parameters, however, as shown by \citet{dawson12}, it is possible to identify highly eccentric Jupiter-sized planets because the LC alone gives a high lower limit on $e$. Therefore, if we are studying Jupiter analogues not very close to their hosts (so that an eccentric orbit cannot be excluded a priori), it is worth following the procedure we have just described. Since  our algorithm produces many files which list all the values assumed by the jump parameters step by step, it is then possible to produce a plot of $e$ versus $\omega$ and compare it with the patterns that are presented in Fig. 2 of \citet{dawson12}. A lower limit $e_{\mathrm{min}}$ on the eccentricity can be set if the observed pattern resembles a pattern for which $g(e,\omega)=\left(\frac{\rho_{\star\mathrm{,circ}}}{\rho_{\star\mathrm{,th}}}\right)^{1/3}\ne1$; that is $\rho_{\star\mathrm{,th}}$ sensibly differs from $\rho_{\star\mathrm{,circ}}$, which is the mean stellar density that would be derived from the transit fit assuming $e=0$. \citet{dawson12} also note that assuming $e=e_{\mathrm{min}}$ implies a transit at periastron (resp. apoastron) if $\rho_{\star\mathrm{,th}}<\rho_{\star\mathrm{,circ}}$ (resp. $\rho_{\star\mathrm{,th}}>\rho_{\star\mathrm{,circ}}$). Thus, launching a second MCMCI run where fixed $e=e_{\mathrm{min}}$ and $\omega=90^{\circ}$ (or 270$^{\circ}$ according to the case) are assumed may facilitate convergence and reduce possible biases in the transit parameters. Of course, awareness that the claimed $e$ value is just a lower limit and that $\omega$ value comes out as a consequence of that must be kept. Instead, if the observed pattern in the $e$-$\omega$ plot is similar to the $g(e,\omega)=1$ scenario, although any $e$ value could be consistent with the transit observable in principle, \citet{dawson12} correctly explain that the planetary orbit likely exhibits a low eccentricity. In this case performing a transit-only fit with the assumption $e=0$ is reasonable.

 In case an $e=0$ orbit is assumed, $\omega$ (which is undefined) is set equal to $0^{\circ}$: the ``periastron'' is placed at the ascending node, thus the time of ``periastron'' is the reference time when RV is maximum if the orbit is circular. Even if exoplanet archives do not have an unambigous convention about the $\omega$ value to adopt when $e=0$, however we call out that the most popular standard that is adopted in the literature is $\omega=90^{\circ}$ \citep[see e.g.][]{eastman13,iglesiasMarzoa15,kreidberg15}.
 
The parameter $K_2$ is used as a jump parameter instead of $K$ to minimise the correlation between jump parameters \citep{gillon09b}. According to its definition, we note that $K_2=(2\pi G)^{1/3} \frac{M_p\sin{i_p}}{M_{\star}^{2/3}}\left(1+\frac{M_p}{M_{\star}} \right)^{-2/3}$, that is it is independent of $e$ and $P$. In particular, for a given stellar mass, assuming a uniform prior on $K_2$ implies assuming a uniform prior on $M_p\sin{i_p}$.

In case of a multi-planetary system, so far only the $\beta$ value of one single planet is handled by the code. We note that $\beta$ can be inferred from the Rossiter-McLaughlin effect and it is not so common to observe this effect on multiple exoplanets belonging to the same system, however we leave the update to multiple values of $\beta$ to a future release of the code.

Stepping in $R_{\star}$ is optional. This option is used for transits too shallow or noisy to make a precise determination of their impact  parameter possible; in this case, $a/R_{\star}$ and $\rho_{\star}$ cannot be well constrained either.

Limb-darkening coefficients can be either four (not-linear model, \emph{nl}) or two (quadratic model, \emph{qd}). In case the \emph{qd} model is used, the assumed jump parameters are not the direct LD coefficients $u_1$ and $u_2$, but the linear combinations $c_1=2u_1+u_2$ and $c_2=u_1-2u_2$ as in \citet{holman06} to minimise the correlation within the uncertainties of LD parameters. For each specified filter, LD coefficients are inferred through proper interpolation in the tables of \citet{claret11}.
Owing to systematics in these tables, we emphasise that simply fixing the LD coefficients at the table values may bias the transit fit or may work backward to bias the $\log{g}$, $T_{\mathrm{eff}}$ or [Fe/H] values from which the LD coefficients are derived. To avoid this scenario, especially in case of high quality photometry, we strongly recommend adding a Bayesian penalty (see later, eq. (\ref{eq:BP})) for the LD coefficients with a prior from the tables. Actually, this is the default behaviour of our code and it has been followed for all the analyses described in the paper.

\subsubsection{Merit function and its role}
The set of input parameters affects both the photometric model reproducing the LC and the RV model reproducing the RV time series (if any); thus at each step we can define
\begin{equation}
 \chi^2_{\mathrm{ph}}=\displaystyle\sum_{i=1}^{n_{\mathrm{LC}}}\sum_{j=1}^{\mathrm{np}_i} \left( \frac{f_{ij}-\bar{f}_{ij}}{\sigma_{f_{ij}}} \right)^2
 \label{eq:photChi2}
,\end{equation}
where n$_{\mathrm{LC}}$ is the number of available LCs, np$_i$ are the number of points of the $i^{\mathrm{th}}$ LC, $f_{i}$, and $\sigma_{f_{i}}$ are the vectors of flux measurements and its errors of the $i^{\mathrm{th}}$ LC, while $\bar{f_i}$ are the fluxes that have been computed according to the eclipse model in use. Adopting the same kind of notation, we can also introduce the analogous $\chi^2_{\mathrm{rv}}$ referring to the RV time series
\begin{equation}
 \chi^2_{\mathrm{rv}}=\displaystyle\sum_{i=1}^{n_{\mathrm{RV}}}\sum_{j=1}^{\mathrm{nr}_i} \left( \frac{v_{ij}-\bar{v}_{ij}}{\sigma_{v_{ij}}} \right)^2
 \label{eq:rvChi2}
.\end{equation}

Summing together $\chi^2_{\mathrm{ph}}$ and $\chi^2_{\mathrm{rv}}$, we obtain the basic merit function. In addition, the jump parameters specified above - besides the stellar luminosity $L_{\star}$, surface gravity $\log{g}$, density $\rho_{\star}$, and projected rotational velocity $v\sin{i_{\star}}$ - may also be set as priors. Those parameters that have been set as Gaussian priors enter the merit function, as well, in the form of the following Bayesian penalty addendum:
\begin{equation}
 \mathrm{BP}=\left( \frac{x-\bar{x}}{\sigma_{\bar{x}}} \right)^2
 \label{eq:BP}
,\end{equation}
where $\bar{x}$ and $\sigma_{\bar{x}}$ are the values with their uncertainties that are attributed to the generic prior parameter, while $x$ comes from the perturbation of $\bar{x}$ at the generic step of the MCMC. In particular, in all the analyses we carried out, we always considered BPs on $c_1$ and $c_2$ (computed from the $u_1$ and $u_2$ LD coefficients). Inference of the initial values for $u_1$ and $u_2$ (with their respective errors) from Claret's tables follows the same procedure as described in \citet{gillon09b}. After that, BP$_{c1}$ and BP$_{c2}$ control the $u_1$ and $u_2$ floating during the MCMC steps, to obtain a LD solution consistent with theory.

Finally, at each step, it is possible to define the complete merit function, which is given by
\begin{equation}
 \chi^2=\chi^2_{\mathrm{ph}}+\chi^2_{\mathrm{rv}}+\displaystyle\sum_{p=1}^{n_{\mathrm{pp}}} \mathrm{BP}_p
 \label{eq:meritChi2}
,\end{equation}
where $n_{\mathrm{pp}}$ is the number of prior parameters. 

Once the priors and the jump parameters have been set, the jump towards the new state is accepted if $\chi^2$ at that step is lower than the one computed at the previous step. If this does not hold, the new state is accepted with probability $\exp{\left(-\frac{1}{2}\Delta\chi^2\right)}$, where $\Delta\chi^2$ is the $\chi^2$ difference between the two last steps. In all the other cases in which the new state is not accepted, the state is pushed back to the previous set of parameters.

Generation of a new state $\mathbf{x'}$ from the present state $\mathbf{x}$ is controlled by a transition probability function and the optimal choice for this function would be considering the posterior probability distribution. However, the aim of the MCMC is actually to retrieve the posterior distribution itself, which is unknown at the beginning. Thus, as a common choice, we considered a Gaussian distribution centred around $\mathbf{x}$. To make the MCMC efficiently converge, the step size, that is the Gaussian variance, to perform the jump from $\mathbf{x}$ to $\mathbf{x'}$ must be tuned up. This is done by computing the acceptance rate, that is the fraction of accepted states over a window spanning a given number of steps, which we set at 100. As reported by \citet{ford05}, if $\mathbf{x}$ has only one dimension, the optimal acceptance rate is $\sim0.44$, while if $\mathbf{x}$ has many dimensions, the optimal acceptance rate is $\sim0.25$, as has been proven by \citet{roberts97}.

We implemented the Gibbs sampling during the burn-in phase, when just a single jump parameter changes at each step; thus, for tuning-up the step size, during the burn-in phase the acceptance rate is set to $40\%$ if the Gibbs sampler is switched on. After the burn-in phase, all the jump parameters are modified at a time, therefore in this phase we monitor the acceptance rate imposing an optimal value equal to $25\%$.

If several sets of models that may differ in terms of input parameters or baseline choice have to be compared, the adopted criterion to choose the best model is given by the Bayesian information criterion \citep{schwarz78}:
\begin{equation}
 \mathrm{BIC}=\hat{\chi}^2+k\log{N}
 \label{eq:BIC}
,\end{equation}
where $\hat{\chi}^2$ is the smallest $\chi^2$ found in the Markov chains, $k$ is the number of free parameters of a given model, and $N$ is the number of data points. The smaller the BIC, the better the model.

At the end, the PDFs of the relevant stellar and planetary parameters are built and meaningful statistics can be retrieved. 

\subsubsection{Convergence discussion}
It is worth launching several chains for a given process and then checking their mutual convergence thanks to the test by \citet{gelman92}, which we briefly present following \citet{ford06}. If $N_c$ is the number of chains, $L_c$ is the length of each chain, and $z_{ic}$ is the $i$-th draw of the generic $z$ parameter at the $c$-th chain, the mean $z$ value within the $c$-th chain is written as
\begin{equation}
 \bar{z}_{\cdot c}=\frac{1}{L_c}\displaystyle\sum_{i=1}^{L_c}z_{ic}
 \label{eq:meanZc}
,\end{equation}
\noindent
while the global mean of $z$ considering the single mean values computed from each chain is written as
\begin{equation}
 \bar{z}_{\cdot\cdot}=\frac{1}{N_c}\displaystyle\sum_{c=1}^{N_c}\bar{z}_{\cdot c}
 \label{eq:meanZ}
.\end{equation}

It is then possible to compute $W(z)$ as the average of the variances within each chain, and $B(z)$ as the variance based on the $\bar{z}_{\cdot c}$ mean values of each chain, that is
\begin{equation}
 W(z)=\frac{1}{N_c}\displaystyle\sum_{c=1}^{N_c}\frac{1}{L_c-1}\displaystyle\sum_{i=1}^{L_c}(z_{ic}-\bar{z}_{\cdot c})^2
 \label{eq:Wvar}
\end{equation}
\begin{equation}
 B(z)=\frac{L_c}{N_c-1}\displaystyle\sum_{c=1}^{N_c}(\bar{z}_{\cdot c}-\bar{z}_{\cdot\cdot})^2
 \label{eq:Bvar}
.\end{equation}
An unbiased estimator of the variance of $z$ is then given by the following weighted average between $W(z)$ and $B(z)$:
\begin{equation}
 \widehat{\mathrm{var}}^+(z)=\frac{L_c-1}{L_c}W(z)+\frac{1}{L_c}B(z)
 \label{eq:unbiasVar}
.\end{equation}

The Gelman-Rubin statistic is commonly denoted by $\hat{R}$, the estimator is defined as
\begin{equation}
 \hat{R}(z)=\sqrt{\frac{\widehat{\mathrm{var}}^+(z)}{W(z)}}
 \label{eq:R_GR}
\end{equation}
\noindent
and convergence is reached when $\hat{R}$ is close to 1. Different threshold values indicating convergence are adopted in the literature: randomly picking up 100 papers that cite \citet{gelman92} in 2017, \citet{vats18} find $\hat{R}$ cut-off values varying from 1.003 to 1.3. Among these papers, $43\%$ of the authors (the relative majority) consider $\hat{R}=1.1$, which is the recommended value by \citet{gelman92}.

Our own empirical practice for convergence is considering $\hat{R}\lesssim1.05$. This criterion may be easily satisfied in case of one single planet fitting, while a slight tension among jump parameters may arise in case of a multi-planetary system. To check convergence, \citet{ford06} also recommends a minimum number of effective independent draws, which can be estimated through
\begin{equation}
 \hat{T}(z)=N_cL_c\min{\left( \frac{\widehat{\mathrm{var}}^+(z)}{B(z)},1 \right)}
 \label{eq:Tdraws}
\end{equation}
with the estimator $\hat{T}(z)\geq1000$.

Since $\hat{R}(z)>1$, from (\ref{eq:R_GR}) we infer
\begin{equation}
 W(z)<\widehat{\mathrm{var}}^+(z).
 \label{eq:Wineq}
\end{equation}
The quantity $\widehat{\mathrm{var}}^+(z)$ comes from an average between $W(z)$ and $B(z)$, therefore its value is between $W(z)$ and $B(z)$; in particular, given condition (\ref{eq:Wineq}), we have
\begin{equation}
 W(z)<\widehat{\mathrm{var}}^+(z)<B(z) \Rightarrow \widehat{\mathrm{var}}^+(z)<B(z)
 \label{eq:Bineq}
.\end{equation}
Condition (\ref{eq:Bineq}) implies
\begin{equation}
 \min{\left(\frac{\widehat{\mathrm{var}}^+(z)}{B(z)},1 \right)}=\frac{\widehat{\mathrm{var}}^+(z)}{B(z)} \Rightarrow \hat{T}(z)=N_cL_c\frac{\widehat{\mathrm{var}}^+(z)}{B(z)}
 \label{eq:Tsimple}
.\end{equation}
By recovering $B(z)$ from eq. (\ref{eq:unbiasVar}) and substituting it in eq. (\ref{eq:Tsimple}), we can express $\hat{T}(z)$ as a function of $\hat{R}(z)$ as follows:
\begin{equation}
 \begin{split}
  \hat{T}(z) & = N_cL_c\frac{\widehat{\mathrm{var}}^+(z)}{L_c\widehat{\mathrm{var}}^+(z)-                  (L_c-1)W(z)} \\
             & = N_cL_c\frac{1}{L_c-(L_c-1)\frac{W(z)}{\widehat{\mathrm{var}}^+(z)}} \\
             & = N_cL_c\frac{1}{L_c-\frac{(L_c-1)}{\hat{R}^2(z)}}
 \end{split}
 \label{eq:T(R)}
,\end{equation}
where we used the definition of $\hat{R}(z)$ given by (\ref{eq:R_GR}) in the last passage. By inverting (\ref{eq:T(R)}), we can finally express $\hat{R}(z)$ as a function of $\hat{T}(z)$, that is
\begin{equation}
 \hat{R}(z)=\sqrt{\frac{L_c-1}{L_c\left(1-\frac{N_c}{\hat{T}(z)} \right)}}
 \label{eq:R(T)}
.\end{equation}
To give some numbers, if we consider $N_c=5$ and $L_c=80'000$,\footnote{It is just a reference value that considers the effective length of a chain, where $20\%$ burn-in length is subtracted to an original chain made of 100'000 steps. Actually eq. (\ref{eq:R(T)}) is almost $L_c$ independent for $L_c$ of order of thousands, which is a common practice in MCMC applications.} achieving the goal suggested by \citet{ford06} ($\hat{T}(z)\ge1000$) would require $\hat{R}(z)\le1.0025$. According to our experience, this $\hat{R}(z)$ cut-off appears too demanding in general. We anticipate that this condition is satisfied for all the jump parameters we considered in the analysis of WASP-4 (\S\ref{ssec:Wasp4} , star orbited by only one hot Jupiter), but it holds just for a few jump parameters of the HD 219134 system (\S\ref{ssec:HD219134}), where the host is orbited by two close-in super Earths (detected both through transit and RV) and by another two, or possibly more, massive planets, whose presence has been inferred from the RV technique. According to all these considerations, we prefer to keep $\hat{R}(z)\lesssim1.05$, as a general reference threshold to establish convergence, and we notice that this cut-off condition is stronger than the average practice in the literature.

\subsubsection{Noise treatment}
As explained in \citet{gillon12}, to estimate the level of noise correlation in each photometric time series, we can define $\beta_w$ as the ratio of the standard deviation of the residuals (observed minus computed (O-C) values) to the mean photometric error: it represents the under- or over-estimation of the white noise. Furthermore, we can take the effect of red noise into account \citep{pont06} by binning the LC on different timescales, and for each binning we then compute
\begin{equation}
 \beta_{r,\mathrm{bin}}=\frac{\sigma_N}{\sigma_1}\sqrt{\frac{N_{\mathrm{bin}}(M-1)}{M}}
 \label{eq:betaRb}
,\end{equation}
where $N_{\mathrm{bin}}$ is the mean number of points in each bin, $M$ the number of bins, and $\sigma_1$ and $\sigma_N$ are the standard deviations of the unbinned and binned residuals, respectively. The maximum among the available $\beta_{r,\mathrm{bin}}$ values is set as the reference $\beta_r$.
Then, the photometric errors provided by the observations are multiplied by the correction factor
\begin{equation}
 \mathrm{CF}=\beta_w\cdot\beta_r
 \label{eq:CF}
\end{equation}
and eventually, for RVs, a ``jitter'' noise may be added quadratically to the error budget. After that a new MCMC run is launched with the updated (rescaled) error bars to obtain reliable error bars on the fitted parameters.

\subsection{Isochrone placement}\label{ssec:IsocPlacement}
The isochrone placement technique aims to infer stellar parameters - above all radius $R_{\star}$, mass $M_{\star}$, and age $t_{\star}$ - by interpolating a set of observational input parameters in a grid of stellar evolutionary models; these observational input parameters come from spectroscopy, transit constraints, photometry and, in particular, they may also involve Gaia parallaxes and magnitudes. The grids of theoretical models in use are currently based upon the PARSEC v.1.2S \citep{chen14,bressan12} + COLIBRI PR16 \citep{marigo13,rosenfield16} stellar isochrones and tracks, which are globally presented in \citet{marigo17} and available for download through the \texttt{CMD} web interface\footnote{\url{http://stev.oapd.inaf.it/cgi-bin/cmd}}.

\subsubsection{Tables of isochrones and evolutionary tracks}

The code is designed so that the stellar location is compared with that of the isochrones, that is loci of evolutionary states occupied by coeval stars.
Each grid of isochrones that has been downloaded from the \texttt{CMD} interface is labelled through the initial metallicity $Z_{\mathrm{ini}}$ of a star at its birth and contains several isochrones. In particular, we selected an age range satisfying the $6<\log{t}<10.1$ (with $t$ in billion years, thereby covering all the plausible stellar age values from pre-MS to post-MS phase) at steps $\delta t=0.05$, which guarantees a fine enough grid considering how slowly stellar parameters evolve and, as a consequence, how large the typical age error bars are. 
Relevant columns reported by the grids are age, mass, effective temperature, luminosity, surface gravity, and absolute magnitudes in a given photometric system; clearly further theoretical parameters as radius or mean stellar density can be easily computed. Summing up, each $Z_{\mathrm{ini}}$ grid of values is made of sub-tables, each of which is representative of an isochrone characterised by a specific age value; these are appended sequentially. Each isochrone is made of several rows and the rows represent theoretical stars of the same age characterised by different parameters, most importantly mass.

As described later in \S\ref{sssec:IntScheme}, isochrone placement also takes into account, for example the stellar evolutionary speed and microscopic diffusion phenomenon, for which $Z$ varies with time. To perform these complementary tasks and extract the needed information, it also needs to deal also with tables of evolutionary tracks. Each track table (labelled through $(M_{\mathrm{ini},\star},Z_{\mathrm{ini}}$) lists the sequence of evolutionary states through which a star having an initial mass $M_{\mathrm{ini},\star}$ and an initial metallicity $Z_{\mathrm{ini}}$ passes during its life. The rows of the table list the evolution of a specific star passing time in terms of, for example mass, radius, and effective temperature. In particular, evolutionary track tables provide the surface abundance of hydrogen and helium at each evolutionary stage, so that the temporal metallicity evolution of a star can be followed, which is essential to face element diffusion (ED).

Considering that the relation between $Z$ and [Fe/H] is exponential, to guarantee linearly spaced [Fe/H] values, it is necessary to take $Z$ values in geometric progression. In particular, to have isochrone grids differing at most by $\delta\mathrm{[Fe/H]}=0.05$ dex, we downloaded all those grids with $Z$ belonging to a geometric progression with common ratio 1.115 from 0.0005 up to 0.00945 and also all those grids corresponding to $Z$ from 0.010 to 0.070 at steps of 0.001. This implies a [Fe/H] range from $-1.48$ dex to $+0.66$ dex.

\subsubsection{Input and output parameters}\label{sssec:inputIsoc}

Isochrone placement was originally implemented in \textsc{MatLab} language as an autonomous and independent routine and then converted into \textsc{Fortran} to be ready to interact with the MCMC code described in \S\ref{ssec:MCMC}; however it is not based on a MCMC procedure. 
In other words, isochrone placement is a fitting algorithm that enables us to retrieve $t_{\star\mathrm{, isoch}}$, $M_{\star\mathrm{, isoch}}$, and $R_{\star\mathrm{, isoch}}$ in agreement with the isochrones, once a useful set of input parameters is provided. 

As a subroutine of the main MCMC, the input parameters of isochrone placement are represented by the relevant set of jump parameters that are generated at each step of the MCMC. 
The role of $M_{\star\mathrm{, isoch}}$ (and possibly $R_{\star\mathrm{, isoch}}$) is to drive the jumps in $M_{\star}$ (and possibly $R_{\star}$) as described in \S\ref{ssec:MCMCI}. At the end, PDFs of $M_{\star}$, $R_{\star}$ and $t_{\star\mathrm{, isoch}}$ are built to extract their respective statistics.

As input parameters, isochrone placement requires the stellar metallicity [Fe/H]$_{\star}$ and necessarily one parameter among stellar effective temperature $T_{\mathrm{eff}}$ and colour index. In addition, it is needed at least one further parameter among distance (observationally available e.g. from parallax) together with apparent magnitude, stellar surface gravity $\log{g}$, stellar mean density $\rho_{\star}$, and stellar radius $R_{\star}$. Optionally one or more parameters among $v\sin{i_{\star}}$, stellar rotational period $P_{\mathrm{rot}}$, $\log{R'_{\mathrm{HK}}}$, and the abundance ratio of yttrium to magnesium [Y/Mg] are useful to better constrain the stellar properties.

In principle, colour index can come from any photometric system supposing that that photometric system is available among the evolutionary models; so far, the code is ready to accept Johnson-Cousin or Gaia photometric systems. 
Stellar surface gravity $\log{g}$ can be available from spectroscopy or via asteroseismology \citep[see e.g. the scaling relation by][which is implemented in our code]{chaplin14}. 
Stellar mean density $\rho_{\star}$ can be observationally available from the LC analysis if a star hosts a transiting planet \citep[see e.g.][]{sozzetti07} or via asteroseismic scaling relations such as that in \citet{chaplin14} that is implemented in our code.
Finally stellar radius $R_{\star}$ may be retrieved from interferometric observations in case of nearby stars.

\subsubsection{Possible removing of isochrone degeneracies}

The role of the $v\sin{i_{\star}}$/$P_{\mathrm{rot}}$, $\log{R'_{\mathrm{HK}}}$ and [Y/Mg] optional parameters is to give a preliminary age indication, that may help in disentangling degeneracies involving MS and pre-MS isochrones, that is isochrone overlap on the Hertzsprung-Russell diagram (HRD). In the case in which both young and old isochrones are geometrically close to the stellar location even a rough age constraint may exclude unlikely very young ages. From each of these optional parameters, an age estimation can be retrieved through empirical relations. It is enough that a MS age is suggested in order to discard unlikely pre-MS isochrones. If this is the case, from a computational point of view, the following interpolation within isochrones considers only those sub-tables of isochrones not containing extremely young ages.

It is well known that rotation and magnetic activity decrease with age as shown by \citet{skumanich72}. The physical explanation originates in the presence of a convective envelope, thus any relation linking age with $P_{\mathrm{rot}}$ (gyrochronology) or $\log{R'_{\mathrm{HK}}}$ (magnetic activity) is to be applied just to late-type stars. This is the typical research domain of the exoplanetary field, so no particular caution need be taken by the user in this respect.

It is also important to stress that gyrochronological and activity-age relations become less sensible to age variations in mid-to-old stars. In fact, as stated by \citet{soderblom10}, it is difficult to detect and calibrate the decline of chromospheric activity in stars older than $\sim2$ Gyr. In addition, \citet{denissenkov10} shows that stellar rotational speeds converge towards similar values after a few billion years, regardless of their initial spin rate; consistently \citet{meibom15} confirm a well-defined period-age relation up to $\sim2.5$ Gyr.
Taking all these considerations into account, \citet{bonfanti16} explain how we use the gyrochronological relation by \citet{barnes10} and the $\log{R'_{\mathrm{HK}}}$-age relation by \citet{mamajek08} to compute conservative age lower limits, which they denote by $\tau_v$ and $\tau_{\mathrm{HK}}$, respectively. Just in assessing age lower limits, the code already takes care of the above-mentioned caveats such that the user is not obliged to be aware of them, and we attain the goal of possible discarding of unlikely pre-MS isochrones before interpolation.

\citet{bonfanti16} also note that the parameter degeneracies on the HRD under the turn-off (TO) can be broken by using $\rho_{\star}$. The mean stellar density is higher in the MS phase than in the pre-MS, thus it enables us to distinguish MS stars from pre-MS stars and to set a conservative age lower limit called $\tau_{\rho}$.

The updated implementation of isochrone placement now also accepts [Y/Mg] among the input parameters. \citet{nissen16} provide the following [Y/Mg]-age relation ($t$ in Gyr; age scatter 0.6 Gyr):
\begin{equation}
 \mathrm{[Y/Mg]}=0.170 - 0.0371 t
 \label{eq:YMg-t}
,\end{equation}
thereby confirming the trend that was first observed by \citet{daSilva12} for solar-type stars.
As explained by \citet{feltzing17} and references therein, yttrium is produced through slow neutron capture process (\emph{s}-process) above all by asymptotic giant branch (AGB) low mass (1-4 $M_{\odot}$) stars. This implies that the interstellar medium is gradually enriched by Y and, as a consequence, younger stars have higher Y abundances.

Eq. (\ref{eq:YMg-t}) holds for solar analogues \citep{nissen16} and also for giant stars of solar metallicity in the helium-core burning phase \citep{slumstrup17}. \citet{feltzing17} show that the [Y/Mg]-$t_{\star}$ trend is also [Fe/H]-dependent and it looses its power of predicting ages for [Fe/H]$\simeq -0.5$ dex, proving that it is only valid for solar metallicity stars. Therefore, if any [Y/Mg] value is provided on input, first of all our code evaluates whether the stellar metallicity belongs to the $-0.2<\mathrm{[Fe/H]}<0.2$ range. If this does not hold, we simply miss the opportunity of removing possible isochronal degeneracies through the [Y/Mg] index, and if neither $v\sin{i_{\star}}$, $\log{R'_{\mathrm{HK}}}$ or $\rho_{\star}$ suggest discarding pre-MS ages, then the isochrone placement scheme would involve the complete set of age values reported by the grids of isochrones. Otherwise, if [Y/Mg] is available and the star has solar metallicity, eq. (\ref{eq:YMg-t}) is considered with the aim of investigating whether the star is old enough to avoid loading pre-MS isochrones. In particular, the grids used for the following interpolation contain all the isochrones that are not younger than the $\tau_{\mathrm{Y}}$ threshold age, which is set as the minimum value between the age computed through eq. (\ref{eq:YMg-t}) and 2.5 Gyr; as for the other ages derived from empirical relations, $\tau_{\mathrm{Y}}$ represents just a conservative age lower limit.

If more than one empirical age is available, among $\tau_v$, $\tau_{\mathrm{HK}}$, $\tau_{\rho}$, and $\tau_{\mathrm{Y}}$, the maximum age indicator is set as the threshold value, such that only evolutionary grids containing higher ages are loaded and used in all the following computation processes to neglect unlikely young ages. We finally observe that it is possible to compute stellar ages from isochrones only if the observational input parameters vary appreciably with time. This does not hold for very low MS stars, which are characterised by very slow evolution. For instance, according to PARSEC evolutionary tracks, a 0.3 $M_{\odot}$ MS star exhibits a $T_{\mathrm{eff}}$ variation $\sim0.1\%$ over the whole age of the universe, which is lower than any reasonable input $T_{\mathrm{eff}}$ error bar.

\subsubsection{Interpolation scheme}\label{sssec:IntScheme}

Loading of isochrones is based upon stellar metallicity; for convenience we note the adopted relation among $Z$ and [Fe/H] \citep[see][]{bonfanti15}, that is
\begin{equation}
 Z=10^{\mathrm{[Fe/H]}-1.817}
 \label{eq:FeH-Zrelation}
\end{equation}
such that [Fe/H]=0 for $Z=Z_{\odot}=0.01524$ consistent with the PARSEC calibration of the Sun \citep{bressan12} taken from \citet{caffau11}. According to (\ref{eq:FeH-Zrelation}), the code initially loads the grid of isochrones corresponding to [Fe/H]$_{\star}$.

If only $T_{\mathrm{eff}}$ and $\rho_{\star}$ (or $\log{g}$) are available as inputs, we work in the modified HRD $\log{T_{\mathrm{eff}}}$-$\log{\rho}$ (or $\log{T_{\mathrm{eff}}}$-$\log{g}$). In all the other combinations of input parameters, it is possible to work in the ordinary HRD, even in the case in which only photometric measurements are available. In fact, if the colour index is among the input parameters, a proper interpolation in the isochrone grid is made to retrieve the corresponding input $T_{\mathrm{eff}}$. Similarly, if both the apparent magnitude and distance are available as inputs, the absolute magnitude is computed and then an interpolation in the grid gives the input stellar luminosity.

If both the gravity proxies $\log{g}$ and $\rho_{\star}$ are available, a rough value for the input stellar mass may be retrieved. Then a preliminary search for at least an evolutionary state that could be contemporarily compatible with the effective temperature, mass, and the two gravity proxies (within their error bars) is performed. In case this preliminary consistency check fails, the input gravity proxy with the largest error bars is discarded to avoid an insolvable tension between mutually dependent input parameters and evolutionary models, which would produce a not satisfactory interpolation.

The star is placed in the HRD and then we take the following steps.

First, further metallic grids of isochrones are loaded compatibly with the [Fe/H] error bars, thereby giving a huge table of isochrones within which we make interpolations (hereafter this table is called \emph{Is}).
 
Second, for each isochrone in each grid, the theoretical evolutionary state that is the closest to the star location in the HRD is considered to then build a new theoretical grid of values (hereafter \emph{Is'}), which is customised according to the stellar location. Theoretical values of these custom points are inferred through linear interpolation considering the two nearest tabulated points. In this way, all the following computations disregard the specific descrete grid of points that were originally tabulated in theoretical models.
 
Third, isochrone points entering the \emph{Is'} customised grid (say $i$ the row index of the grid) are smoothed through a bi-dimensional Gaussian window function
 \begin{equation}
  \mathcal{G}(y_i, T_{\mathrm{eff},i})= \frac{e^{-\frac{1}{2}\left[\left(\frac{\log{T_{\mathrm{eff},i}}-\log{T_{\mathrm{eff}}}}{\Delta\log{T_{\mathrm{eff}}}}\right)^2+
  \left(\frac{\log{y_{\star,i}}-\log{y_{\star}}}{\Delta\log{y_{\star}}}\right)^2\right]}}
  {2\pi \Delta\log{T_{\mathrm{eff}}} \Delta\log{y_{\star}}}
  \label{eq:Gauss}
 ,\end{equation}
 where $y_{\star}$ may indicate either $\log{L_{\star}}$, $\log{\rho_{\star}}$, or $\log{g}$, depending on the specific HRD we are working in.
 
 Fourth, we define the weight
 \begin{multline}
  p_{i}=\left[\left(\frac{\log{T_\mathrm{eff}}-\log{T_{\mathrm{eff,}i}}}{\Delta\log{T_{\mathrm{eff}}}}\right)^2+
  \left(\frac{y_{\star}-y_{\star,i}}{\Delta y_{\star}}\right)^2+
  \sum_j \left(\frac{\tilde{y}_{j\star}-\tilde{y}_{j\star,i}}{\Delta \tilde{y}_{j\star}}\right)^2 \right.\\
  \left. +\log^2{\left(\frac{v_{\mathrm{ref}}}{v_{\mathrm{evo}}}\right)}\right]^{-1}
  \label{eq:peso}
 \end{multline}
 to be attributed to the $i$-th row of \emph{Is'}, where $y_{\star}$ has the same meaning as at the previous point, while $\tilde{y}_{j\star}$ may represent, if directly available, the input $M_{\star}$ and $\log{g}$ (being $\tilde{y}_{j\star} \neq y_{\star} \forall j$). The weight takes into account both the geometrical distance between theoretical and observational parameters and the evolutionary speed $v_{\mathrm{evo}}$ of a star in that particular region of the HRD, that is
 \begin{equation}
  v_{\mathrm{evo}}=\sqrt{\left(\frac{y_2-y_1}{t_2-t_1}\right)^2+\left(\frac{\log{T_{\mathrm{eff,2}}}-\log{T_{\mathrm{eff,1}}}}{t_2-t_1}\right)^2}
 \label{eq:vevo}
 ,\end{equation}
 where ($\log{T_{\mathrm{eff,1}}}$, $y_1$) and ($\log{T_{\mathrm{eff,2}}}$, $y_2$) are two consecutive points with ages $t_1$ and $t_2$ along a track displaying the evolutionary path of a theoretical star having the same mass and metallicity of the star to be analysed. The quantity $v_{\mathrm{ref}}$ is a normalisation factor corresponding to the minimum evolutionary speed that has been detected along the whole track.
 
 The less $v_{\mathrm{evo}}$, the greater the probability of finding a star in that evolutionary stage. According to Eq. (\ref{eq:peso}), the weight of a theoretical point increases as the distance ``theory-observations'' decreases and $v_{\mathrm{evo}}$ decreases.
 
 Finally, the generic $X_{\star}$ parameter and its uncertainty $\Delta X_{\star}$ to be retrieved according to stellar evolutionary models are then computed through
 \begin{equation}
  X_{\star}=\frac{\displaystyle\sum_{i=1}^{\bar{n}}{X_{i}\mathcal{G}(L_i, T_{\mathrm{eff},i})p_i}}{\displaystyle\sum_{i=1}^{\bar{n}}{\mathcal{G}(L_i, T_{\mathrm{eff},i})p_i}}.
  \label{eq:X.is}
 \end{equation}
 \begin{equation}
  \Delta X_{\star}=\sqrt{\frac{\displaystyle\sum_{i=1}^{\bar{n}}{(X_{i}-X_{\star})^2\mathcal{G}(L_i, T_{\mathrm{eff},i})p_i}}{\displaystyle\sum_{i=1}^{\bar{n}}{\mathcal{G}(L_i, T_{\mathrm{eff},i})p_i}}},
  \label{eq:DeltaX.is}
 \end{equation}
 where $\bar{n}$ is the number of points (rows) of the grid \emph{Is'}. 
 
At this point the ED phenomenon, also known as microscopic or atomic diffusion \citep{burgers69,chapman70,chaboyer01}, enters the picture of the algorithm. Element diffusion is caused by the interaction among various chemical species in stars with a convective envelope, which leads to a progressive surface depletion of elements heavier than hydrogen, which sink downwards. 

In particular, it turns out that very low mass stars up to 0.5 $M_{\odot}$ exhibit constant $Z$ during their entire life; this is also the case for stars more massive than $\sim2$ $M_{\odot}$. In the intermediate mass range, instead, $Z$ diminishes during the MS phase. After that, as the convective envelope deepens, the consequent mixing of elements makes the star approximately as metallic as it was at its birth \citep[see e.g.][Fig.~3]{bonfanti15}.

As already said, the isochrone grids are identified through $Z_{\mathrm{ini}}$, while we know only the present day metallicity of a star $Z_{\star}$. 
As shown by \citet[Fig.~4]{bonfanti15}, neglecting the ED effect may result in older age derivations, especially for intermediate age stars. Also \citet{dotter17} observed that when initial metallicity is equated to the current metallicity, the resulting isochronal ages of stars can be systematically overestimated by up to $\sim20\%$. 
Assessing a wrong evolutionary state results in bad mass and radius estimation as well. For this reason, starting from evolutionary tracks (which tabulate $Z$ values along time), we built up a piecewise third degree polynomial interpolation of the curves $Z_{k,l}=Z_{k,l}(t)$, for any given stellar mass and $Z_{\mathrm{ini}}$ (represented through the subscripts $l$ and $k$, respectively). In this way, after an iterative process, it is possible to infer what the $Z_{\mathrm{ini}}$ of a given star was, starting from its present $Z_{\star}$ and a first raw guess of its age and mass. In fact, at the first iteration, the isochrone grid identified by the present day metallicity of the star is employed to derive a first guess for stellar age ($t_{1,\star}$) and mass ($M_{1,\star}$).
After computing $Z_{k,M_{1,\star}}(t=t_{1,\star})$ for all the available curves characterised by different $Z_{\mathrm{ini}}$ (identified by the subscript $k$), we identify the $\hat{k}^{\mathrm{th}}$ curve such that $Z_{\hat{k},M_{1,\star}}(t=t_{1,\star})=Z_{\star}$, that is the first guess for the initial metallicity of the star $Z_{1,\mathrm{ini,}\star}$.

By loading the grid of isochrones labelled $Z_{1,\mathrm{ini,}\star}$ and repeating all the algorithms described above for retrieving theoretical stellar parameters, new output values for stellar age and mass are obtained. The entire ED procedure is repeated iteratively until a convergence in the $Z_{\mathrm{ini},\star}$ values is reached.

According to what we have just described, our isochrone placement evaluates many model stars at each chain step and it emerges with the stellar properties that best match the rest of the parameters. Since an optimisation methodology within each step may lead to underestimated uncertainties, we decided to compare the age uncertainties inferred from the $t_{\star\mathrm{,isoch}}$ PDFs of HD 219134 and WASP-4 (see our analysis and results in \S\ref{sec:results}) to that stated in the literature. From the review by \citet{soderblom10}, who evaluates the works by \citet{pont04}, \citet{jorgensen05}, and \citet{takeda07} in assessing isochronal ages, it turns out that typical isochronal age uncertainties are in the range $\sim20\%$-$50\%$ (before considering systematics on evolutionary models). Age uncertainties we provide for our test stars are $\sim36\%$, $\sim19\%$, and $\sim34\%$, thus they are consistent with what specialists in the field usually report.
We also note that our ages do not purely come from isochrones, and in fact activity index and gyrochronology are used to better constrain them. In this regard, \citet{angus19} point out that the combination of isochrone fitting and gyrochronology improves age precision, thus age uncertainties that are lower than those coming from the isochronal analysis alone are expected and our claimed formal uncertainties do not appear underestimated.
Therefore, our specific implementation of the optimisation algorithm does not seem to lead to underestimated uncertainties, and we also stress the importance of adding systematic uncertainties (see \S\ref{ssec:isoPrec}) to our formal uncertainties. However, an optimisation algorithm may be a potential source of underestimated uncertainties and letting the MCMC sample the isochrones directly \citep[like in ][]{eastman19} represents a valid alternative to our implementation scheme.

\subsection[MCMCI]{MCMCI\footnote{The MCMCI code is available at \url{https://github.com/Bonfanti88/MCMCI}}}\label{ssec:MCMCI}

If in its original implementation the MCMC required a specification of a type of priors or relation involving $M_{\star}$ and $R_{\star}$ to fully characterise the exoplanetary system, the current MCMCI input form enables us to select the use of theoretical models. Isochrone placement, which is called as a subroutine by the main MCMC, is in charge of deriving $M_{\star}$ and $R_{\star}$ (besides the stellar age $t_{\star}$) as a consequence of the stellar input parameters to be specified, as explained in \S\ref{sssec:inputIsoc}. 

The step parameters always adopted on the stellar side are comprised of [Fe/H], $T_{\mathrm{eff}}$ (or the colour index), and $M_{\star}$, where [Fe/H] and $T_{\mathrm{eff}}$ (or the colour index) are also set as priors. Since [Fe/H] is always present among the prior parameters that are randomly perturbed at each step of the chains and several chains of order $\sim10^5$ steps are expected to be launched, it would be very time consuming to load the specific metallic grids of isochrones and tracks that are expected to vary from step to step each time. Therefore we decided to load a reasonable amount of evolutionary models since the beginning of the algorithm and to store these in very big matrices. By keeping track of initial and ending indices identifying the various models in the matrices, at each step only the needed sub-matrices are actually used in computations.

Two basic situations are possible in which we specifically draw attention to what is essential to retrieve $M_{\star}$ and $R_{\star}$ thanks to evolutionary models.

In the first case scenario the transit is very deep and well defined, such that its impact parameter $b$ can be precisely measured from the LC. We step in [Fe/H], $T_{\mathrm{eff}}$, $M_{\star}$, and $W$.
 From $W_{\mathrm{step}}$, $\frac{a}{R_{\star}}_{\mathrm{step}}$ is computed through eq. (\ref{eq:aR}). From Kepler's third law, $\frac{a}{R_{\star}}_{\mathrm{step}}$ is translated into $\rho_{\star\mathrm{,step}}$, and $R_{\star\mathrm{,step}}$ is derived from $M_{\star\mathrm{,step}}$ and $\rho_{\star\mathrm{,step}}$. In addition, $\rho_{\star\mathrm{,step}}$ enters isochrone placement (together with [Fe/H]$_{\mathrm{step}}$ and $T_{\mathrm{eff,step}}$,) to derive the theoretical $M_{\star\mathrm{, isoch}}$ at that step, according to eq. (\ref{eq:X.is}). Finally, the Bayesian penalty
 \begin{equation}
  \mathrm{BP_M}=\left( \frac{M_{\star\mathrm{, step}}-M_{\star\mathrm{, isoch}}}{\Delta M_{\star}} \right)^2
  \label{eq:BP_M}
 \end{equation}
 (where $\Delta M_{\star}$ is the uncertainty of $M_{\star\mathrm{, isoch}}$, that is computed according to eq. (\ref{eq:DeltaX.is})) is added to the merit function given by eq. (\ref{eq:meritChi2}). In this way, $M_{\star\mathrm{,step}}$ is driven by the theoretical mass value $M_{\star\mathrm{,isoch}}$, and thus $R_{\star\mathrm{,step}}$ is computed from $M_{\star\mathrm{,step}}$ (constrained by evolutionary models) and $\rho_{\star\mathrm{,step}}$ (constrained by the transit);
 
 In the second case scenario the transit is very shallow, such that $b$ is not directly inferrable from the LC. Besides [Fe/H], $T_{\mathrm{eff}}$ and $M_{\star}$, $R_{\star}$ is a also jump parameter in this case and may be also set as prior. At each step, input parameters for isochrone placement are [Fe/H]$_{\mathrm{step}}$, $T_{\mathrm{eff,step}}$, and at least one indirect piece of information involving the stellar radius (e.g. the stellar luminosity through apparent magnitude and parallax, the spectroscopic $\log{g}$, or an interferometric measurement of the radius). Then, according to eqs. (\ref{eq:X.is}) and (\ref{eq:DeltaX.is}), isochrone placement gives $M_{\star\mathrm{, isoch}}\pm\Delta M_{\star}$ and $R_{\star\mathrm{, isoch}}\pm\Delta R_{\star}$ as output values, which enter the following Bayesian penalties:
 \begin{equation}
  \mathrm{BP_{M}}+\mathrm{BP_{R}}=\left( \frac{M_{\star\mathrm{, step}}-M_{\star\mathrm{, isoch}}}{\Delta M_{\star}} \right)^2 + \left( \frac{R_{\star\mathrm{, step}}-R_{\star\mathrm{, isoch}}}{\Delta R_{\star}} \right)^2
  \label{eq:BP_MR}
 \end{equation}
 to be added to the merit function expressed by eq. (\ref{eq:meritChi2}). The quantity $\rho_{\star\mathrm{,step}}$ is computed consistently with $M_{\star\mathrm{,step}}$ and $R_{\star\mathrm{,step}}$, whose values are driven by the theoretical $M_{\star\mathrm{, isoch}}$ and $R_{\star\mathrm{, isoch}}$ coming out at each step. Thanks to $\rho_{\star\mathrm{,step}}$, $\frac{a}{R_{\star}}_{\mathrm{step}}$ is computed through Kepler's third law and finally $W_{\mathrm{step}}$ is expressed as a function of $\frac{a}{R_{\star}}_{\mathrm{step}}$ by inverting eq. (\ref{eq:aR}), such that transit parameters are constrained a posteriori by evolutionary models. We note that if the stellar radius has been also set as a prior as a result of the BP given by eq. (\ref{eq:BP}), its value also feeds back into $\rho_{\star}$ and into the corresponding transit parameters in general.

From $M_{\star\mathrm{,step}}$ and $R_{\star\mathrm{,step}}$, all the relevant planetary parameters can be computed, especially $R_{p\mathrm{,step}}$ (from $\mathrm{d}F_{\mathrm{step}}$, if any LC is available) and $M_{p\mathrm{,step}}$ (if any RV time series or, at least, $K$ is available). At the end of all the chains, PDFs of the parameters of the exoplanetary system are built upon the step values. 

After a first MCMCI run when CFs referring to each LC are computed through eq. (\ref{eq:CF}), the photometric error bars are properly rescaled to obtain reliable error bars on the fitted parameters. If everything else is the same, the MCMCI is launched a second time to retrieve reliable PDFs of both stellar and planetary parameters.

Algorithm tests and all the subsequent analysis that is presented in \S\ref{sec:results} have been run on a \textsc{Dell} Latitude laptop (1 TB SSD; 16 GB RAM; Intel core i7-7820 HQ @ 2.90 GHz; FSB=3.68 GHz). One single run for analysing HD 219134 system (4 LC time series for a total amount of 3322 data points, 1 RV time series made of 663 data points, four orbiting planets, element diffusion switched-on) took $\sim30$ hours. One single run for analysing Wasp-4 system (14 LC time series for a total amount of 4010 data points, one orbiting planet, element diffusion switched-on) took $\sim20$ hours. If we considered a very basic fit of one single planet orbiting a star with just 1 available LC, the expected run-time is seven to eight hours if ED is switched on.

By switching off ED in performing the isochrone placement technique, we save a factor of 2 in run-time. However, as we already stressed, neglecting ED may imply an age overestimation up to $\sim20\%$. Instead, the impact on the isochronal $M_{\star}$ and $R_{\star}$ can be estimated up to a few percent.

We are aware of other implementations of the MCMC algorithm, such as the differential evolution \citep[DE;][]{terBraak06} or the affine invariant \citep[AI;][]{foreman13} technique. We are currently working on the implementation of the AI-MCMC procedure, which will also contain Gaussian processes \citep{rasmussen05} for modelling correlated noise, and according to our first estimations, this technique will increase the run-time speed by a factor 10. The full \textsc{Fortran} implementation of the AI-MCMC requires likely one year, since it is necessary to revise the architecture of our current MCMC code. Thus, we prefer to share our current MCMCI version now and then considering the ongoing update as part of a future work.

\section{Analysis and results}\label{sec:results}
To test our algorithm we selected two planet-hosting stars that are representative of two different test cases. One of these stars is HD 219134, hosting several low mass planets, which has already been studied by, for example \citet{motalebi15}, \citet{vogt15}, and \citet{gillon17}. The other is WASP-4, which is known to host a heavily bloated hot Jupiter owing to transit analyses already carried out by \citet{wilson08}, \citet{gillon09a}, \citet{winn09}, \citet{southworth09}, \citet{sanchisOjeda11}, and \citet{nikolov12}.

\subsection{HD 219134}\label{ssec:HD219134}
HD 219134 (also known as GJ 892 or HIP 114622) is a bright close-by K3 V star. Its relevant input parameters used during our analyses are listed in Tab. \ref{tab:HD219134}. Our first attempt to analyse this system took the tabulated $T_{\mathrm{eff}}$ among input parameters, but no convergence was reached in theoretical models.

\begin{table}
\caption{HD 219134. Input stellar parameters considered in this work}
\label{tab:HD219134}
\centering
\begin{tabular}{l l l l}
\hline\hline       
\multicolumn{2}{c}{Parameters}  & HD 219134 & \multicolumn{1}{c}{References} \\ 
\hline                    
 [Fe/H]             & [dex]     & $0.11\pm0.04$     & \citet{motalebi15}  \\  
 $T_{\mathrm{eff}}$ & [K]       & $4699\pm16$       & \citet{boyajian12}  \\
 $v\sin{i_{\star}}$ & [km/s]    & $0.4\pm0.5$       & \citet{motalebi15}  \\
 $R$                & [$R_{\odot}$]& $0.778\pm0.005$& \citet{boyajian12}  \\
 $V$                & [mag]     & 5.57              & \citet{vanLeeuwen07}\\
 $B-V$              & [mag]     & 0.99              & \citet{vanLeeuwen07}\\
 $d$                & [pc]      & $6.5325\pm0.0038$ & \citet{GaiaDR2_18} \\
 $\log{R'_{\mathrm{HK}}}$&      & $-5.02\pm0.06$    & \citet{motalebi15}  \\
 $P_{\mathrm{rot}}$ & [days]    & $42.3\pm0.1$      & \citet{motalebi15}  \\
\hline
\end{tabular}
\end{table}

\begin{figure}
 \centering
 \includegraphics[width=\columnwidth]{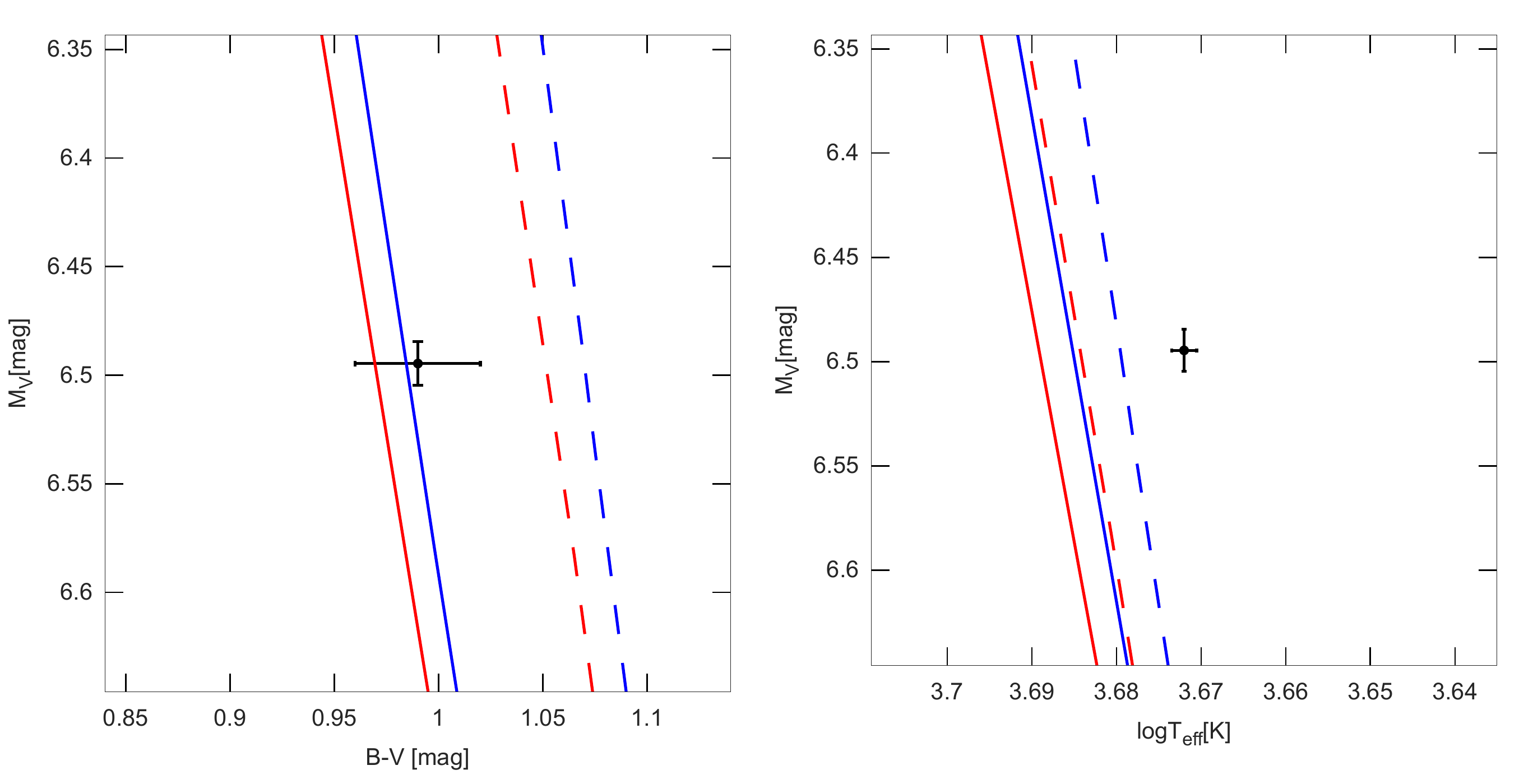}
 \caption{HD 219134 location on the colour-magnitude diagram (left panel) and on the $\log{T_{\mathrm{eff}}}$-$M_V$ diagram (right panel) together with some reference isochrones. Red stands for 1 Gyr models, while blue stands for 10 Gyr models; solid lines are representative of the nominal stellar metallicity ($Z=0.020\rightarrow$ [Fe/H]=0.11); dashed lines refer to extremely super-solar metallicity $Z=0.070\rightarrow$ [Fe/H]=0.66.}
 \label{fig:CMDvsTMD}
\end{figure}

\begin{figure}
 \centering
 \includegraphics[width=\columnwidth]{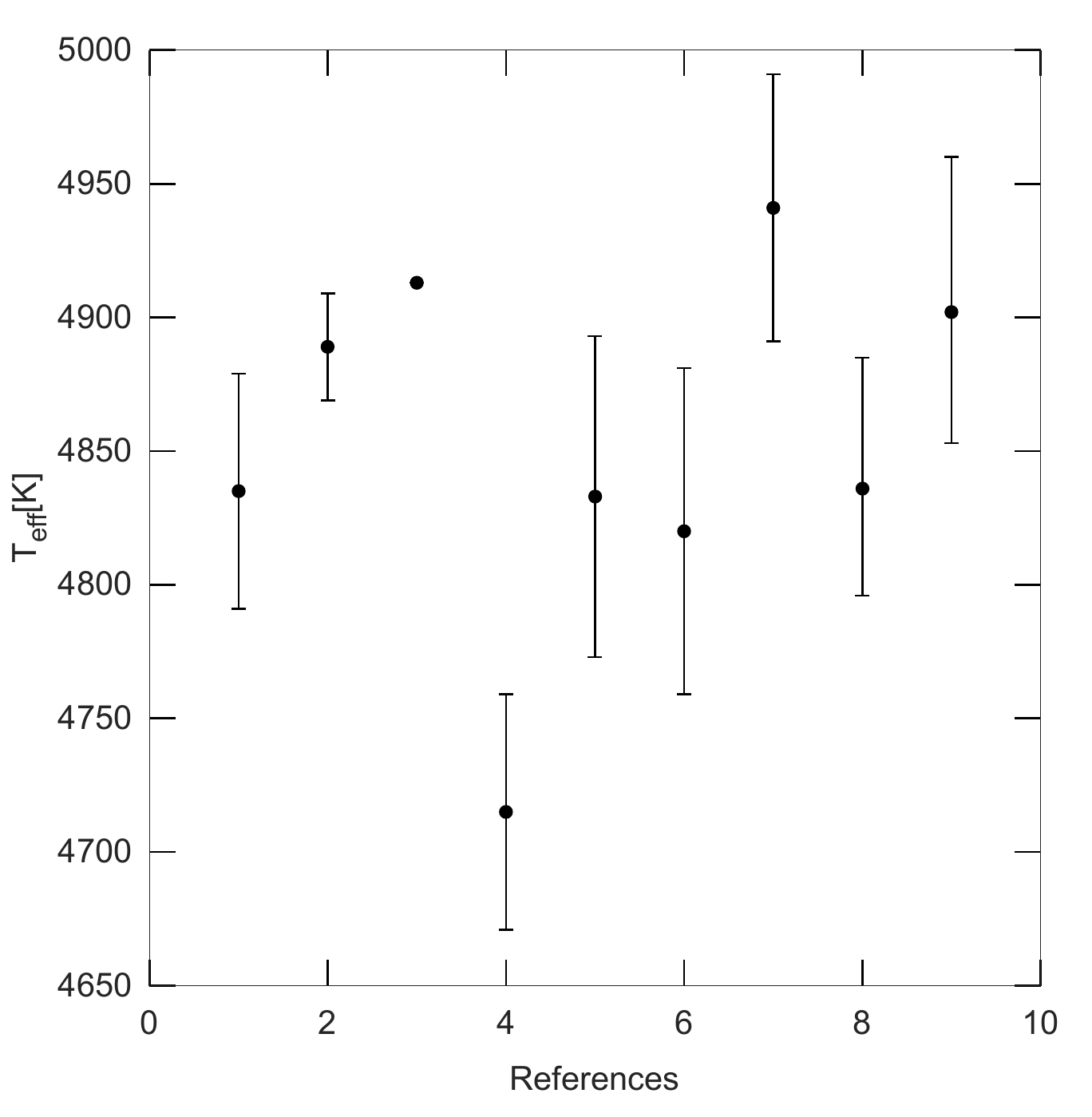}
 \caption{Values of $T_{\mathrm{eff}}$ found in the literature for HD 219134. Values on the $x$-axis correspond to the following references: 1) \citet{valenti05}; 2) \citet{mishenina08}; 3) \citet{soubiran08}; 4) \citet{prugniel11}; 5) \citet{ramirez13}; 6) \citet{motalebi15}, EW approach; 7) \citet{motalebi15}, SPC approach; 8) our work, $L$ approach; and 9) our work, $R$ approach.}
 \label{fig:HD219134Teff}
\end{figure}

Actually, as shown in Fig. \ref{fig:CMDvsTMD} (right panel), the $T_{\mathrm{eff}}$ value provided by \citet{boyajian12}, which has been also considered, for example by \citet{motalebi15} and \citet{gillon17}, is completely inconsistent with theoretical models, regardless of the metallicity. This value was empirically computed by \citet{boyajian12}, considering the interferometric measurement of the stellar diameter (which yields $R=0.778R_{\odot}$) and an empirical estimate of the bolometric flux by fitting photometry to spectral templates. On the other hand we observed that the $B-V$ colour index (which is more straightforward to obtain) has a reasonable value, if compared with stellar models (Fig. \ref{fig:CMDvsTMD}, left panel). In addition, interpolation inside theoretical models suggests corresponding $T_{\mathrm{eff}}=4836$ K or 4902 K (depending on the $L$- or $R$-approach; see later), which agree with other estimates provided in the literature as shown in Fig. \ref{fig:HD219134Teff}. For these reasons, we decided to start from $B-V$, rather than $T_{\mathrm{eff}}$.

The transit analysis was carried out because of the four LCs observed through the Spitzer/IRAC detector \citep{fazio04} and the RV time series gathered by the HARPS-N spectrograph \citep{cosentino12} already considered in \citet{gillon17}. Both the nominal model made of one star and four planets and the choice of the baseline functions were taken from \citet{gillon17}. We recall that two LCs contain the transit of planet b, while the other two the transit of planet c. Planets d and f are not transiting and their presence was deduced from the RV analysis. 

We notice that there are claims of further orbiting planets in the literature. \citet{motalebi15} talk about four planets containing the transit of b, c, d, plus one long period planet that orbits the star with $P=1842$ days, but without considering planet f of our analysis. Instead, \citet{vogt15} claim the presence of six planets; besides b, c, d, and f which we considered, they also find planet g with $P=94.2$ days and planet h with period $P=2247$ days. The high polynomial degree (order 4) that enters the temporal de-trending of the RV time series suggests the likely presence of other (long period) planets besides the four we selected. On the other hand, there is not full agreement in the literature about the total number and period of the further planets, and deeply investigating and characterising the presence of other planets is beyond the scope of this paper; instead, we want to show the behaviour of our MCMCI in front of a challenging multi-planetary system as HD 219134.

For all four planets, $T_0$, $P$, and $K_2$ were assumed as jump parameters. Also the eccentricity $e$ has been let free to vary, except for planet b, for which a fixed $e=0$ value was set, according to \citet{gillon17}, who did not register any improvement in the BIC by let $e$ vary. Consistently, \citet{gillon17} state that the closiness of HD 219134 b to its star implies a circularisation timescale of 80 Myr even assuming as tidal quality factor the maximum value that is derived for terrestrial planets and satellites of the Solar System \citep{murray99}. The same computation for planet c yields a timescale of $2.5$ Gyr, therefore we preferred letting $e$ free to jump for this and the other planets.
In case of the two transiting planets b and c as well as $\mathrm{d}F$ and $b'$ were set as jump parameters. The transits result to be too shallow to reliably constrain $W$. We preferred not to set this a priori, but rather to count on observational stellar parameters and on theoretical evolutionary models to characterise the star completely so that $W$ was later inferred. 

On the stellar side, two different sets of input parameters were considered. The first set considers {[Fe/H]}, $B-V$, $P_{\mathrm{rot}}$, $\log{R'_{\mathrm{HK}}}$, $V$ magnitude and parallactic distance $d$. Interpolation within isochrones enabled us to recover input stellar luminosity $L$ from $V$ and $d$; Bayesian penalties (BPs) involving [Fe/H], $B-V$ and $L$ were added to the merit function by setting the corresponding parameters as priors. Hereafter the analysis starting from this input set is called Luminosity prior approach ($L$ approach).

The second set considers {[Fe/H]}, $B-V$, $P_{\mathrm{rot}}$, $\log{R'_{\mathrm{HK}}}$, input stellar radius $R_{\mathrm{prior}}$ from interferometry. In this case BPs entering the merit function involved [Fe/H], $B-V$ and $R_{\mathrm{prior}}$. Hereafter the analysis starting from this input set is called Radius prior approach ($R$ approach).

The MCMCI was launched considering five chains of 100'000 steps each, for both the $L$ and $R$ approach. In both cases, a \emph{qd} model for LD was assumed and $c_1$ and $c_2$ were properly computed to build their corresponding BPs to be added to the merit function. Furthermore, also the sum $\mathrm{BP_{M}} + \mathrm{BP_{M}}$ given by (\ref{eq:BP_MR}) was added to the merit function.

\begin{figure}
 \includegraphics[width=\columnwidth]{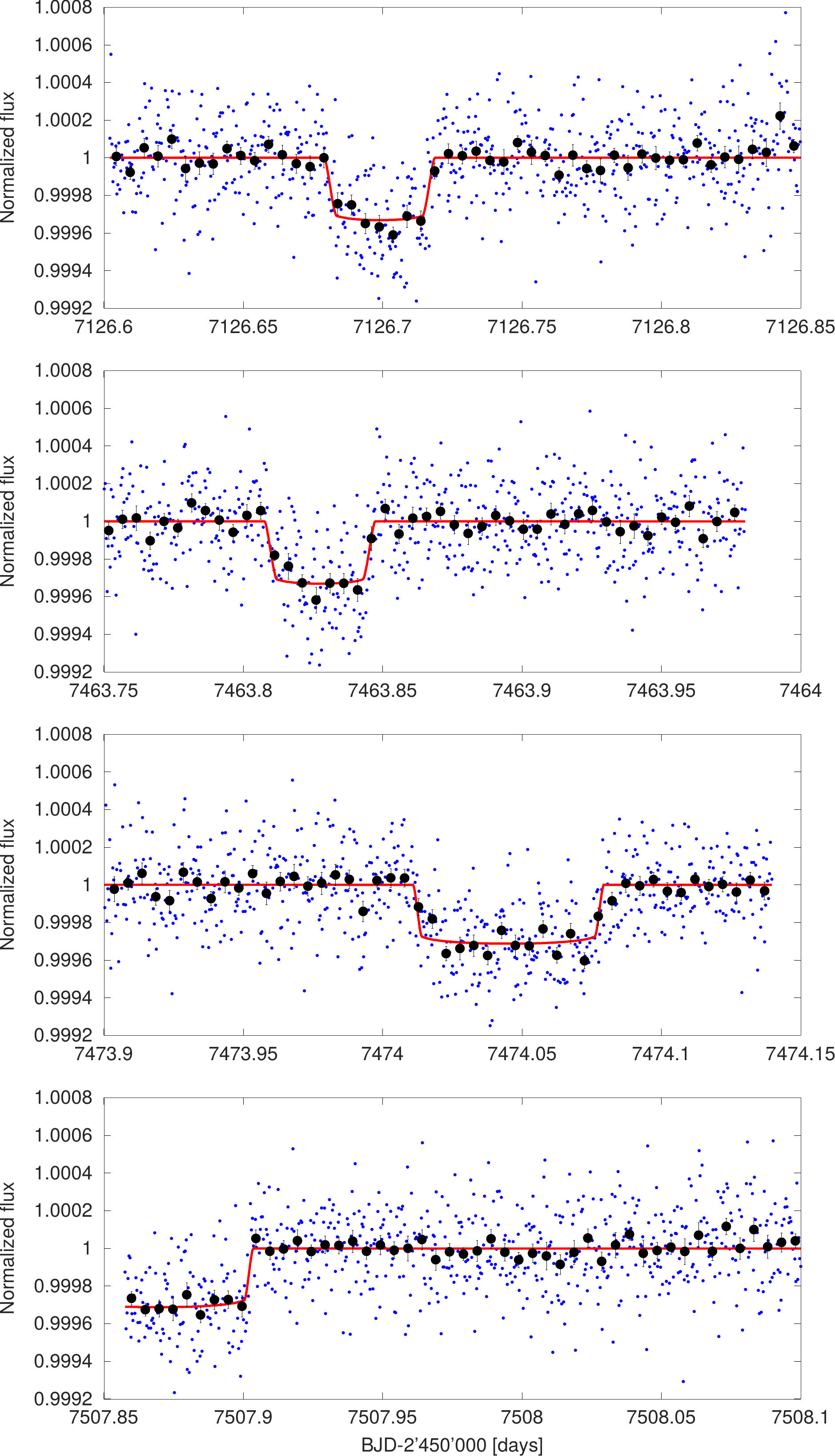}
 \caption{Superimposed transiting models on LCs, as derived from the $L$ approach; the $R$ approach gives analogous results. The two top panels refer to the transit of HD 219134 b, while the two bottom panels refer to HD 219134 c.}
 \label{fig:HD219134LC}
\end{figure}

After a first run, the MCMCI was launched again by applying the CFs recovered at the first run through eq. (\ref{eq:CF}), to retrieve all the relevant stellar and planetary parameters (convergence checked thanks to the Gelman-Rubin test), that are summarized in Tab. \ref{tab:HD219134output} and \ref{tab:HD219134planets}. In particular, the transiting HD 219134 b and c are confirmed to be two super-Earth rocky planets. Their corresponding LCs with the super-imposed transit models as inferred from the $L$-approach are displayed in Fig. \ref{fig:HD219134LC}.

\begin{table}
\caption{HD 219134. Output stellar parameters derived considering $L$ approach or $R$ approach}
\label{tab:HD219134output}
\centering
\begin{tabular}{l l l l}
\hline\hline       
\multicolumn{2}{c}{Parameters}  & \multicolumn{1}{c}{$L$ approach} & \multicolumn{1}{c}{$R$ approach} \\ 
\hline                    
 $T_{\mathrm{eff}}$ & [K]       & $4836^{+49}_{-40}$             & $4902^{+58}_{-49}$    \\
 $M_{\star}$        & [$M_{\odot}$]& $0.788^{+0.017}_{-0.016}$   & $0.799^{+0.023}_{-0.019}$\\
 $R_{\star}$        & [$R_{\odot}$]& $0.7508^{+0.0095}_{-0.0099}$&$0.7774\pm0.0050$         \\
 $\rho_{\star}$     & [$\rho_{\odot}$]& $1.854^{+0.12}_{-0.080}$   &$1.790^{+0.057}_{-0.046}$\\
 $\log{g}$          & [cgs]     & $4.576^{+0.021}_{-0.014}$      & $4.553^{+0.013}_{-0.011}$ \\
 $L_{\star}$        & [$L_{\odot}$]& $0.2772^{+0.0048}_{-0.0056}$& $0.313^{+0.016}_{-0.014}$         \\
 $t_{\star}$        & [Gyr]     & $8.7^{+1.8}_{-4.5}$            & $9.8^{+0.85}_{-2.9}$     \\
\hline
\end{tabular}
\end{table}

\begin{table*}
\caption{Planets hosted by HD 219134. Top portion shows the results coming from $L$ approach, while bottom portion shows the results coming from $R$ approach.}             
\label{tab:HD219134planets}      
\centering          
\begin{tabular}{l l l l l l l }
\hline\hline       
\multicolumn{3}{c}{Planets} & \multicolumn{1}{c}{b} & \multicolumn{1}{c}{c} & \multicolumn{1}{c}{d} & \multicolumn{1}{c}{f} \\ 
\hline                    
 Eclipse timing & $T_0$ & [BJD]\tablefootmark{(a)} & $7126.69912^{+0.00090}_{-0.00087}$ & $7474.04589^{+0.0010}_{-0.00081}$ & $7735.48^{+0.84}_{-0.51}$ & $7716.35^{+0.47}_{-0.52}$ \\  
 Orbital period   & $P$   & [days]  & $3.092926^{+0.000011}_{-0.000010}$ & $6.76457^{+0.00028}_{-0.00034}$ & $46.854\pm0.027$ & $22.719^{+0.012}_{-0.015}$ \\
 Semi-major axis  & $a$   & [AU]   & $0.03839^{+0.00028}_{-0.00027}$ & $0.06468^{+0.00047}_{-0.00045}$ & $0.2350^{+0.0017}_{-0.0016}$ & $0.1450\pm0.0010$ \\
 Irradiation      & $S$   & [$S_{\oplus}$]& $188.1\pm3.8$ & $66.2\pm1.3$ & $5.02\pm0.10$ & $13.17^{+0.27}_{-0.26}$ \\
 Equilibrium temperature\tablefootmark{(b)} & $T_{\mathrm{eq}}$ & [K]& $1032.0\pm5.2$ & $795.0\pm4.0$ & $417.1\pm2.1$ & $530.9\pm2.7$ \\
 Transit depth    & $\mathrm{d}F$  & [ppm]         & $356^{+24}_{-25}$ & $317\pm19$ &  & \\
 Impact parameter & $b$  & [$R_{\star}$] & $0.9203^{+0.0068}_{-0.0070}$ & $0.845^{+0.024}_{-0.019}$ &  & \\
 Transit duration & $W$   & [min]         & $56.4^{+1.6}_{-1.7}$ & $99.4^{+2.4}_{-2.2}$ & & \\
 Orbital inclination & $i_p$ & [$^{\circ}$]& $85.185^{+0.13}_{-0.087}$ & $87.382^{+0.095}_{-0.090}$ &  &  \\
 Orbital eccentricity& $e$ &              & 0 (fixed) & $0.060^{+0.043}_{-0.036}$ & $0.114^{+0.033}_{-0.020}$ & $0.157\pm0.047$ \\
 Argument of pericentre\tablefootmark{(c)} & $\omega$ & [$^{\circ}$] & 0 (fixed) & $70^{+23}_{-39}$ & $175^{+12}_{-11}$ & $83^{+16}_{-17}$ \\
 RV semiamplitude   & $K$ & [m/s]         & $2.389^{+0.067}_{-0.070}$ & $1.692^{+0.078}_{-0.073}$ & $3.31^{+0.11}_{-0.10}$ & $1.910^{+0.093}_{-0.11}$ \\
 Mass\tablefootmark{(d)}  & $M_p$ & $M_{\oplus}$& $4.62\pm0.14$ & $4.23^{+0.20}_{-0.19}$ & $15.67^{+0.53}_{-0.51}$ & $7.06^{+0.34}_{-0.40}$ \\
 Radius             & $R_p$ & $R_{\oplus}$& $1.544^{+0.056}_{-0.059}$ & $1.458^{+0.047}_{-0.048}$ & & \\
 Density            & $\rho_p$ &[$\rho_{\oplus}$]& $1.25^{+0.16}_{-0.13}$ & $1.36^{+0.16}_{-0.14}$ &  & \\
\hline
 Eclipse timing & $T_0$ & [BJD]\tablefootmark{(a)} & $7126.69913^{+0.00092}_{-0.00082}$ & $7474.04590^{+0.0010}_{-0.00078}$ & $7735.32^{+0.42}_{-0.38}$ & $7716.36^{+0.45}_{-0.47}$ \\  
 Orbital period   & $P$   & [days]        & $3.0929259\pm0.0000093$ & $6.76456^{+0.00029}_{-0.00036}$& $46.851^{+0.028}_{-0.027}$ & $22.718\pm0.014$ \\
 Semi-major axis  & $a$   & [AU]          & $0.03856^{+0.00037}_{-0.00030}$ & $0.06496^{+0.00062}_{-0.00051}$ & $0.2360^{+0.0022}_{-0.0018}$ & $0.1457^{+0.0014}_{-0.0011}$ \\
 Irradiation      & $S$   & [$S_{\oplus}$]& $210.8^{+8.9}_{-8.3}$ & $74.2^{+3.1}_{-2.9}$ & $5.62^{+0.24}_{-0.22}$ & $14.76^{+0.62}_{-0.58}$ \\
 Equilibrium temperature\tablefootmark{(b)} & $T_{\mathrm{eq}}$ & [K]& $1062\pm11$ & $818.0^{+8.5}_{-8.2}$ & $429.1^{+4.4}_{-4.3}$ & $546.2^{+5.7}_{-5.4}$ \\
 Transit depth    & $\mathrm{d}F$  & [ppm]         & $356^{+23}_{-22}$ & $317\pm19$ &  & \\
 Impact parameter & $b$  & [$R_{\star}$] & $0.9253^{+0.0052}_{-0.0050}$ & $0.859^{+0.025}_{-0.024}$&  & \\
 Transit duration & $W$   & [min]         & $56.7^{+1.2}_{-1.4}$ & $99.8^{+2.7}_{-2.3}$&  & \\
 Orbital inclination & $i_p$ & [$^{\circ}$]& $85.020^{+0.067}_{-0.062}$ & $87.260^{+0.087}_{-0.085}$ &  & \\
 Orbital eccentricity& $e$ &              & 0 (fixed) & $0.061^{+0.039}_{-0.038}$ & $0.110^{+0.016}_{-0.019}$ & $0.159^{+0.048}_{-0.051}$\\
 Argument of pericentre\tablefootmark{(c)} & $\omega$ & [$^{\circ}$] & 0 (fixed) & $70^{+24}_{-38}$ & $175^{+12}_{-13}$ & $84\pm16$ \\
 RV semiamplitude   & $K$ & [m/s]         & $2.393^{+0.076}_{-0.073}$ & $1.692^{+0.076}_{-0.077}$ & $3.30\pm0.10$ & $1.896\pm0.099$ \\
 Mass\tablefootmark{(d)}  & $M_p$ & $M_{\oplus}$& $4.68\pm0.17$ & $4.27\pm0.21$ & $15.80\pm0.56$ & $7.07^{+0.40}_{-0.38}$\\
 Radius             & $R_p$ & $R_{\oplus}$& $1.601^{+0.052}_{-0.051}$ & $1.511\pm0.046$ &  & \\
 Density            & $\rho_p$ &[$\rho_{\oplus}$]& $1.14^{+0.12}_{-0.11}$ & $1.24^{+0.14}_{-0.12}$ &  & \\

\hline                  
\end{tabular}
\tablefoot{\tablefoottext{a}{TDB time standard, shifted by $-2'450'000$. More precisely $T_0$ refers to the time of conjunction, as already specified in the text; }
\tablefoottext{b}{Assuming zero albedo; }
\tablefoottext{c}{Placed at the ascending node in case of circular orbit;}
\tablefoottext{d}{Actually $M_p\sin{i_p}$ for planets d and f.}
}
\end{table*}

A comparison involving radii and masses of HD 219134 and its planets as derived by various authors is presented in Tab. \ref{tab:HD219134cmp}. \citet{gillon17} analyse the system assuming stellar $R_{\mathrm{prior}}$ as prior similarly to our $R$ approach and find that very good agreement holds between these two estimations. From Tab. \ref{tab:HD219134cmp} we also notice that our uncertainties affecting $M_p$ and $R_p$ are generally lower than those reported in the literature, especially in the case of planetary masses. This suggests that the integration of stellar theoretical models with the MCMC algorithm is able to reduce the PDF width of the output parameters.

\begin{table*}
\caption{Comparison involving masses and radii derived by different authors for HD 219134 system.}
\label{tab:HD219134cmp}
\centering
\begin{tabular}{cllllll}
\hline\hline
 \multicolumn{2}{c}{HD 219134} & \multicolumn{1}{c}{$L$ approach} & \multicolumn{1}{c}{$R$ approach} & \multicolumn{1}{c}{(1)} & \multicolumn{1}{c}{(2)} & \multicolumn{1}{c}{(3)} \\
\hline
\multirow{2}*{Star} & $R_{\star}$ [$R_{\odot}$] & $0.7508^{+0.0095}_{-0.0099}$&$0.7774\pm0.0050$  & $0.778\pm0.005$ & $0.778\pm0.005$ & $0.77\pm0.02$ \\
& $M_{\star}$ [$M_{\odot}$] & $0.788^{+0.017}_{-0.016}$   & $0.799^{+0.023}_{-0.019}$ & $0.81\pm0.03$ & $0.78\pm0.02$ & $0.794^{+0.037}_{-0.022}$ \\
\hline
Planet b & $R_p$ [$R_{\oplus}$] & $1.544^{+0.056}_{-0.059}$ & $1.601^{+0.052}_{-0.051}$ & $1.602\pm0.055$ & $1.606\pm0.086$ & \\
$P=3.1$ days & $M_p$ [$M_{\oplus}$] & $4.62\pm0.14$ & $4.68\pm0.17$ & $4.74\pm0.19$ & $4.36\pm0.44$ & $3.8\pm0.3$ \\
\hline
Planet c & $R_p$ [$R_{\oplus}$] & $1.458^{+0.047}_{-0.048}$ & $1.511\pm0.046$ & $1.511\pm0.047$ &  & \\
$P=6.8$ days & $M_p$ [$M_{\oplus}$] & $4.23^{+0.20}_{-0.19}$ & $4.27\pm0.21$ & $4.36\pm0.22$ & $2.78\pm0.65$\tablefootmark{(c)} & $3.5\pm0.6$ \\
\hline
Planet d & \multirow{2}*{$M_p\sin{i_p}$ [$M_{\oplus}$]} & \multirow{2}*{$15.67^{+0.53}_{-0.51}$} & \multirow{2}*{$15.80\pm0.56$} & \multirow{2}*{$16.17\pm0.64$} & \multirow{2}*{$8.94\pm1.13$}\tablefootmark{(c)} & \multirow{2}*{$21.3\pm1.3$} \\
$P=46.9$ days\tablefootmark{(a)} &  &  &  &  &  &  \\
\hline
Planet f & \multirow{2}*{$M_p\sin{i_p}$ [$M_{\oplus}$]} & \multirow{2}*{$7.06^{+0.34}_{-0.40}$} & \multirow{2}*{$7.07^{+0.40}_{-0.38}$} & \multirow{2}*{$7.30\pm0.40$} &  & \multirow{2}*{$8.9\pm1.0$} \\
$P=22.7$ days\tablefootmark{(b)} &  &  &  &  &  &  \\
\hline
\end{tabular}
\tablefoot{(1): \citet{gillon17}; (2): \citet{motalebi15}; (3): \citet{vogt15} \\
\tablefoottext{a}{Both (2) and (3) detected $P=46.7$ days. }
\tablefoottext{b}{(3) detected $P=22.8$ days}
\tablefoottext{c}{Reported as the minimum mass}
}
\end{table*}

Finally, we comment on the results we obtained with our two different and independent $L$ and $R$ approaches with a particular reference to our derived $R_{\star}$ values. 
Input $V$ magnitude and Gaia parallax are totally independent of (and possibly inconsistent with) the interferometric measure of the radius. Actually, the comparison between the $L$ and $R$ approach aims to investigate the mutual consistency of input parameters by looking at the output results.

\begin{figure}
 \includegraphics[width=\columnwidth]{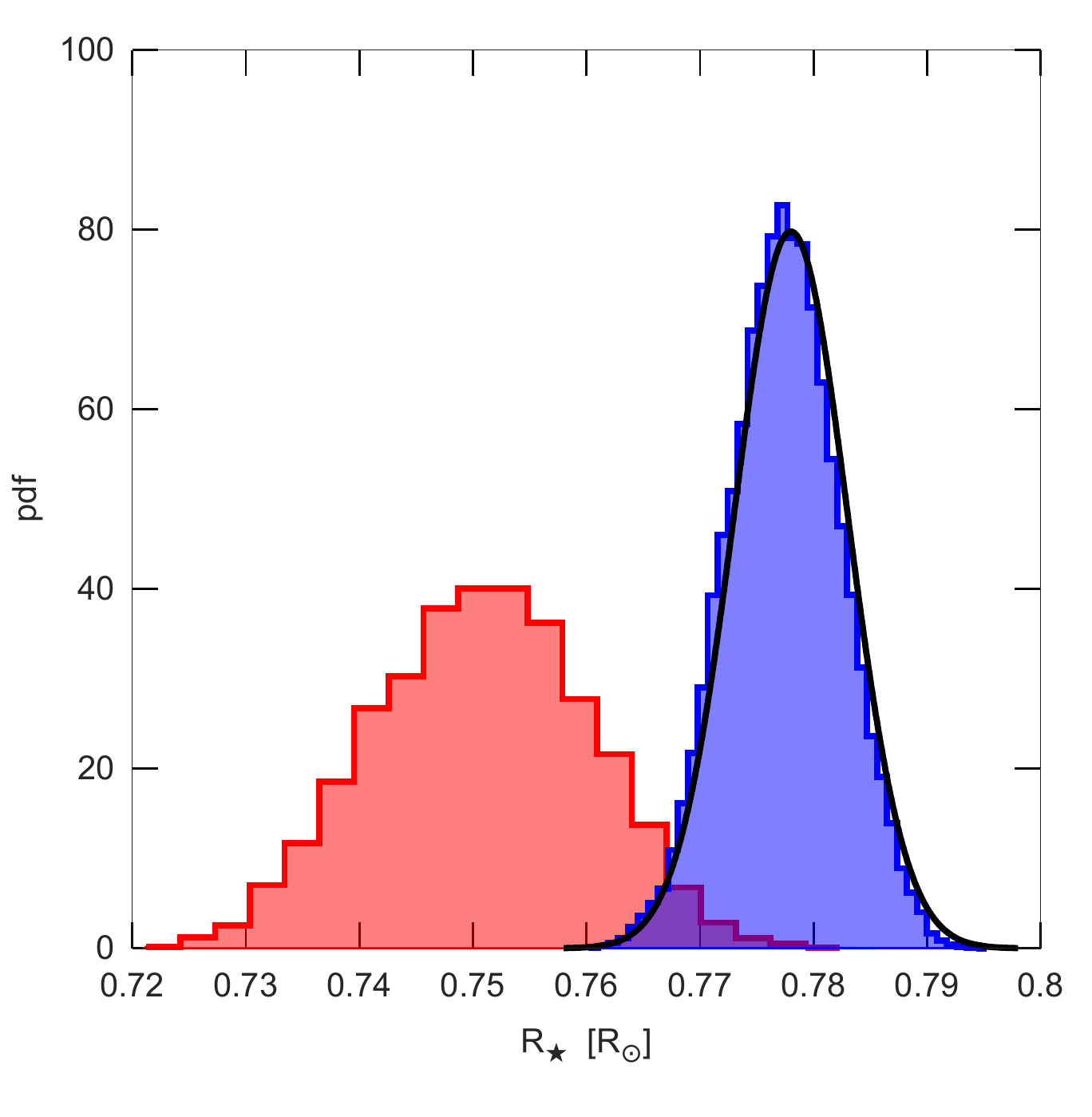}
 \caption{The red histogram is the posterior $R_{\star}$ PDF, as derived from the $L$ approach. The blue histrogram is the posterior $R_{\star}$ PDF, as from the $R$ approach. The black Gaussian is the interferometric prior on stellar radius that has been imposed in the $R$ approach. The black Gaussian prior essentially overrides the isochronal constraint (which is looser) and drives the output $R_{\star}$ value. See text for a complete discussion.}
 \label{fig:Rdistrib}
\end{figure}

In Fig. \ref{fig:Rdistrib} the red histogram is the output $R_{\star}$ PDF, which has been inferred considering ([Fe/H]$_{\mathrm{step}}$, $B-V_{\mathrm{step}}$, $L_{\star,{\mathrm{step}}}$) as inputs for isochrone fitting and assuming a prior on the parallax-based stellar luminosity $L_{\mathrm{prior}}$. Instead, the blue histogram is the output $R_{\star}$ PDF that has been inferred considering ([Fe/H]$_{\mathrm{step}}$, $B-V_{\mathrm{step}}$, $R_{\star,{\mathrm{step}}}$) as inputs for isochrone fitting, and assuming a prior (black Gaussian) on the interferometric radius $R_{\mathrm{prior}}$.

It is important to stress that the red histogram is the isochronal-consistent $R_{\star}$ PDF when $L_{\star\mathrm{,step}}$ enters the fitting algorithm. On the other hand, if $R_{\star\mathrm{,step}}$ is assumed to be an input for isochrone placement instead of $L_{\star\mathrm{,step}}$, the isochronal-consistent $R_{\star}$ PDF that is produced in output is expected to be consistent with $R_{\star\mathrm{,step}}$, and thus possibly shifted with respect to the red histogram if the input radius and luminosity are not consistent for the given $T_{\mathrm{eff}}$ and [Fe/H]. Actually, if $R_{\star\mathrm{step}}$ is assumed on input, the isochrone-based Bayesian penalty $\mathrm{BP_R}$ alone (eq. (\ref{eq:BP_MR})) would drive to an output radius $R_{\mathrm{out,isoch}}\approx0.767\pm0.020R_{\odot}$: if compared with the $L$ approach, the amplified uncertainty suggests that isochrones better agree with the luminosity input values, rather than the radius input values. If we now consider that the $R$ approach is also characterised by the presence of the additional black Gaussian prior whose $\sigma=0.005 R_{\odot}$, it turns out that its correspondent BP (eq. (\ref{eq:BP})) plays the most prominent role in determining the output $R_{\star}$ PDF. The blue histogram enters this scenario consistently, as it is just slightly shifted towards lower values with respect to the black Gaussian.

The inconsistence of the derived $R_{\star}$ values (red versus blue histogram in Fig. \ref{fig:Rdistrib}) is the consequence of the inconsistence between the parallax-based stellar luminosity and the interferometric radius, for the given photometry and metallicity. Reasons for this tension may be found in the input photometry and [Fe/H], surface gravity that is implied by stellar position in the HRD (which all influence $T_{\mathrm{eff}}$ and bolometric correction, upon which $L$ is inferred), intrinsic limits of evolutionary models, and the interferometric technique for retrieving stellar radius, which may suffer some systematics \citep[see e.g.][]{white18}.



\subsection{WASP-4}\label{ssec:Wasp4}
WASP-4 is a $V=12.5$ mag G7V star orbited by a transiting hot Jupiter. We carried out the transit analysis with two LCs by \citet{winn09} (Sloan $z'$ band exposures taken at Las Campanas Observatory in Chile using the MagIC camera) and 12 LCs derived from three transits simultaneously observed in the four Sloan $g'$, $r'$, $i'$, $z'$ bands by \citet{nikolov12} at La Silla Observatory. We are aware of the availability of other LCs since other studies of WASP-4 have been already done, as specified at the beginning of \S\ref{sec:results}. Carrying out an analysis considering all the available material in the literature is beyond the scope of the paper, which instead tests the MCMCI and its ability to contemporaneously deal with several LCs taken in different bands.

By launching multiple small runs of the MCMCI (chains of 10'000 steps each) and monitoring the BIC value after varying the baseline, we realised that the photometric time series do not need a baseline function more complex than a simple scalar (normalisation). We considered $\mathrm{d}F$, $b'$, $W$, $T_0$, and $P$ as jump parameters and, in addition, we set $T_{\mathrm{eff}}$ and [Fe/H] as priors. Thus BPs on $T_{\mathrm{eff}}$ and [Fe/H] were considered, besides those on the $c_1$ and $c_2$ LD coefficients (\emph{qd} model assumed). Moreover, $\mathrm{BP_M}$ from eq. (\ref{eq:BP_M}) was also added to the merit function. The orbit was assumed to be circular, following the indications by \citet{sanchisOjeda11}; these authors report that no eccentricity has been revealed in the study of several RV time series \citep{wilson08,madhusudhan09,pont11} and occulation data \citep{beerer11}. We complemented this information by specifying also $v\sin{i_{\star}}$ to improve the convergence within theoretical models and the RV semi-amplitude $K$ to get an output value for the planetary mass.

\citet{bouma19} has stressed that the sequence of transit recently observed by TESS happened earlier than expected. The authors speculate that this timing variation may be caused by tidal orbital decay or apsidal precession, while it seems unlikely to be the presence of a third body that could perturb the star-planet system. Anyway, we decided to launch a run in which transit timing variations \citep[TTVs;][]{holman05} were also allowed in our model, but the BIC we obtained was higher with respect to the case in which TTVs were not considered. Therefore we avoided including TTVs in our analysis.

As in the case of HD 219134, first the MCMCI code was launched to estimate the CFs (eq. (\ref{eq:CF})) and then it was launched a second time after applying the CFs. Five chains of 100'000 steps each were considered at each run and the convergence of the derived parameters was checked using the Gelman-Rubin test. As a reference, the two LCs by \citet{winn09} and the three LCs observed in the $g'$ band by \citet{nikolov12} are shown in Fig. \ref{fig:Wasp4LC}, together with the superimposed transit models.

Input and derived parameters of the WASP-4 system are listed in Tab. \ref{tab:Wasp4}. We note that the relative uncertainty on $\rho_{\star}$ is lower than those affecting $M_{\star}$ and $R_{\star}$. In fact, unlike the case of HD 219134, in this work $\rho_{\star}$ is constrained from the transit\footnote{Constraining stellar density from transit with a relative uncertainty lower than $2\%$ as in this case is not so common, and it has been possible because of the prominent transit depth and high quality LCs. According to NASA Exoplanet Archive, the median uncertainty for transit-inferred $\rho_{\star}$ is $\delta\rho_{\star,t}\sim9\%$, and only $\sim5\%$ of the systems have $\delta\rho_{\star,t}<2\%$.} and then used to derive $(M_{\star},R_{\star})$, which depend on $\rho_{\star}$ itself and on the evolutionary stellar models. Various considerations hold for $\log{g_{\star}}$ and $\rho_p$: they are computed from the pairs $(M_{\star},R_{\star})$ and $(M_p,R_p)$, respectively, thus according to error propagation, we could expect their relative uncertainties to be higher than those affecting $M_{\star}$, $R_{\star}$, $M_p$, and $R_p$. 
But this is not the case since we have a derived output uncertainty $\left.\frac{\Delta g_{\star}}{g_{\star}}\right|_{\mathrm{out}}=0.020$ versus $\left.\frac{\Delta g_{\star}}{g_{\star}}\right|_{\mathrm{ep}}=0.063$ that should be expected from error propagation; similarly, we have $\left.\frac{\Delta \rho_p}{\rho_p}\right|_{\mathrm{out}}=0.026$ versus $\left.\frac{\Delta \rho_p}{\rho_p}\right|_{\mathrm{ep}}=0.065$.

\begin{figure}
 \includegraphics[width=\columnwidth]{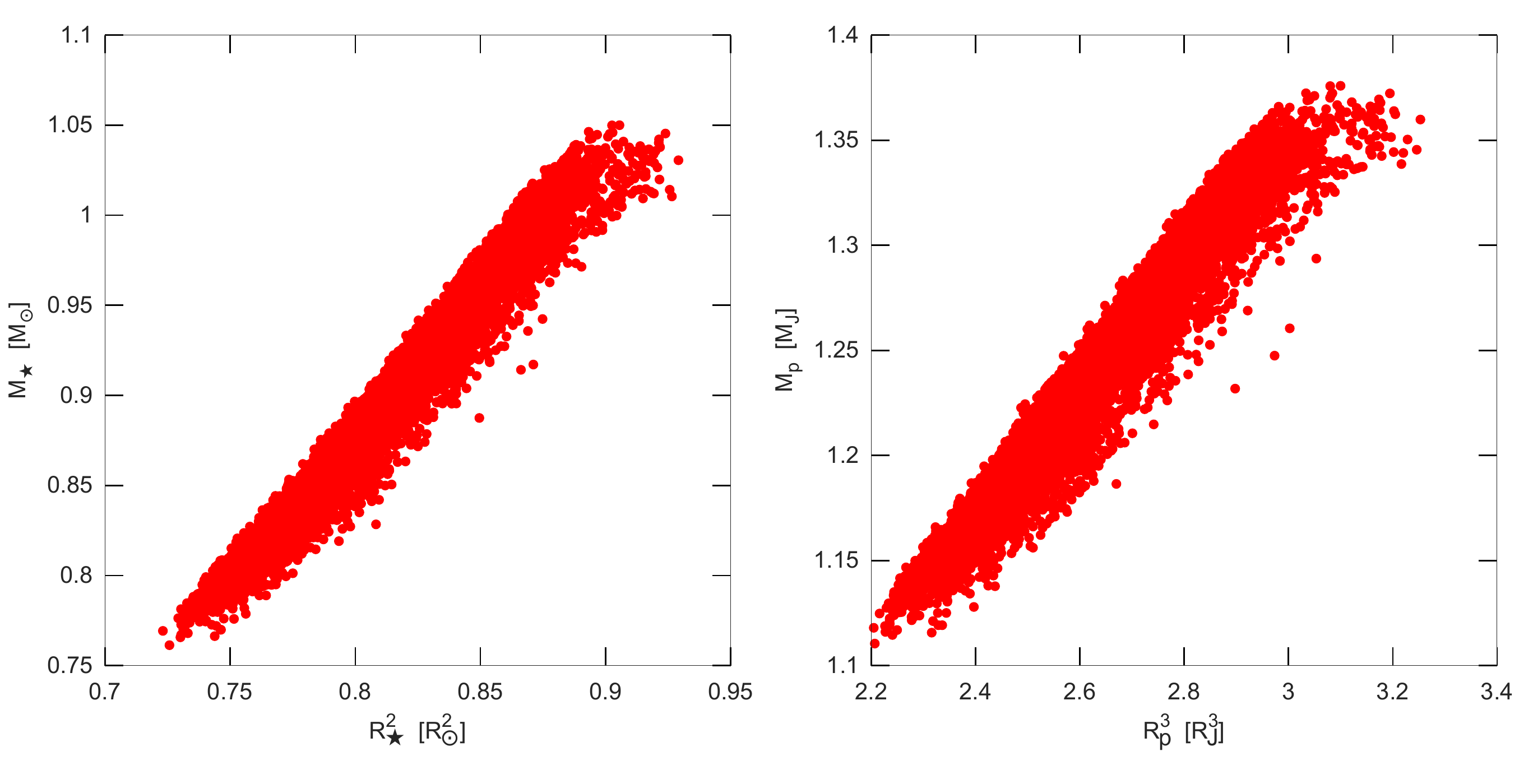}
 \caption{WASP-4: regression of $M_{\star}$ vs. $R_{\star}^2$ (left panel), and of $M_p$ vs. $R_p^3$ (right panel). A strong correlation is expected because $\rho_{\star}$ is constrained from the transit, and $(M_p,R_p)$ are related to $(M_{\star},R_{\star})$, transit parameters and the RV semi-amplitude.}
 \label{fig:MsRsMpRp}
\end{figure}

\begin{figure}
 \includegraphics[width=\columnwidth]{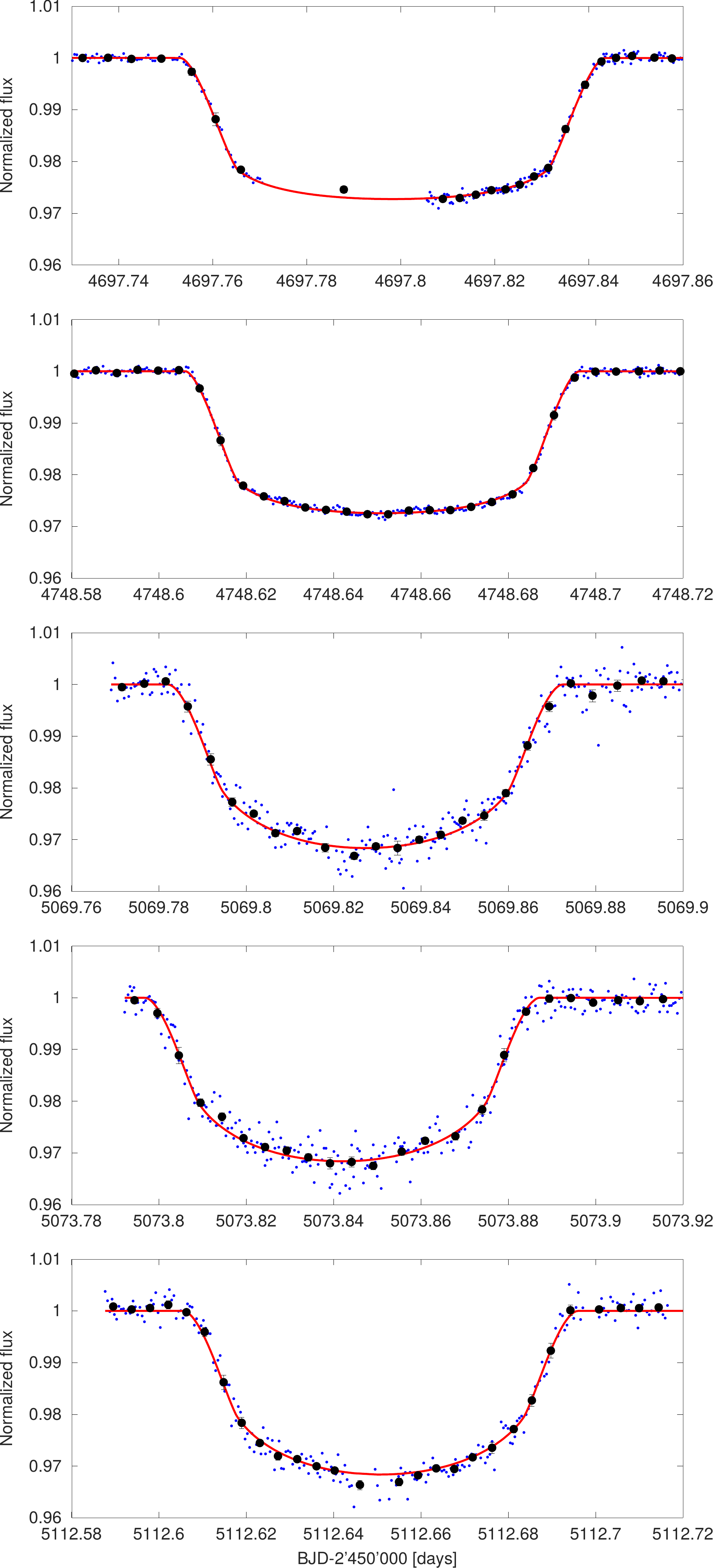}
 \caption{Light curves of WASP-4 b. First two panels show the data provided by \citet{winn09}, who observed the transit in the Sloan $z'$ band. The other three panels show the observation by \citet{nikolov12} in the Sloan $g'$ band. The red thick line represents the transit model as inferred from our analysis.}
 \label{fig:Wasp4LC}
\end{figure}

However, uncertainties in agreement with standard error propagation are expected only if the pairs $(M_{\star},R_{\star})$ and $(M_p,R_p)$ are uncorrelated. Actually, a correlation between $M_{\star}$ and $R_{\star}$ exists because, besides isochrones, we have a strong constraint on $\rho_{\star}\sim\frac{M_{\star}}{R_{\star}^3}$ thanks to the transit. In addition, $M_p$ (resp. $R_p$) differ from $M_{\star}$ (resp. $R_{\star}$) by factors that are constrained from the transit and the RV semi-amplitude; thus a correlation between $M_p$ and $R_p$ is also expected, although it is a bit more diluted. What has just been described is shown in Fig. \ref{fig:MsRsMpRp}, where the $M_{\star}$ (resp. $M_p$) values coming from all the chain steps are plotted versus the corresponding $R_{\star}^2$ (resp. $R_p^3$) values. Linear correlations are clear and the coefficients of determination are $r^2_{\star}=0.91$ (left panel) and $r^2_p=0.87$ (right panel).

If a strict linear correlation held ($r^2=1$), no dispersion in $\log{g_{\star}}\sim\frac{M_{\star}}{R_{\star}^2}$ and in $\rho_p\sim\frac{M_p}{R_p^3}$ would be registered, thus both their uncertainties would be zero. The equation $\sigma_{\mathrm{ne}}=\sqrt{1-r^2}$ represents the fraction of root mean square that is not explained by the linear regression, thus it is statistically expected that error bar extensions coming from simple error propagation (where the involved quantities are considered independent) may be reduced up to a factor $\sigma_{\mathrm{ne}}$ or so. Statistically, uncertainties may go down to the following estimated levels:
\begin{equation}
 \left.\frac{\Delta g_{\star}}{g_{\star}}\right|_{\mathrm{exp}}=\left.\frac{\Delta g_{\star}}{g_{\star}}\right|_{\mathrm{ep}} \cdot \sqrt{1-r^2_{\star}} = 0.019 \approx \left.\frac{\Delta g_{\star}}{g_{\star}}\right|_{\mathrm{out}}
 \label{eq:Dg_g}
\end{equation}
\begin{equation}
 \left.\frac{\Delta \rho_p}{\rho_p}\right|_{\mathrm{exp}}=\left.\frac{\Delta \rho_p}{\rho_p}\right|_{\mathrm{ep}} \cdot \sqrt{1-r^2_p} = 0.024 \approx \left.\frac{\Delta \rho_p}{\rho_p}\right|_{\mathrm{out}}
 \label{eq:Drho_rho}
\end{equation}
It turned out that the statistically expected uncertainties (indicated by the subscript ``exp'') are in agreement with the actual derived ones (indicated by the subscript ``out'').

Tab. \ref{tab:Wasp4cmp} reports a comparison involving stellar and planetary masses and radii, as derived by different authors. Our results have the advantage of retrieving $M_{\star}$ and $R_{\star}$ owing to the strict interaction of the MCMC algorithm with the stellar evolutionary models, which yields to very precise $M_p$ and $R_p$ values, if compared to what is stated in the literature.

\begin{table}
\caption{WASP-4 planetary system. Top portion: input stellar parameters as from \citet{gillon09a}; middle portion: derived stellar parameters; bottom portion: planetary parameters.}
\label{tab:Wasp4}
\centering      
\begin{tabular}{l l l}
\hline\hline          
\multicolumn{3}{c}{WASP-4} \\
\hline
 [Fe/H] & [dex] & $-0.03\pm0.09$ \\
 $T_{\mathrm{eff}}$ & [K] & $5500\pm100$    \\
 $v\sin{i_{\star}}$ & [km/s] & $2\pm1$     \\
 $K$                & [m/s]     & $247.6^{+13.9}_{-6.8}$     \\
\hline
 $M_{\star}$        & [$M_{\odot}$]& $0.898^{+0.048}_{-0.046}$ \\
 $R_{\star}$        & [$R_{\odot}$]& $0.902\pm0.016$ \\
 $\rho_{\star}$     & [$\rho_{\odot}$]& $1.225^{+0.015}_{-0.022}$ \\
 $\log{g}$          & [cgs]     & $4.4745^{+0.0086}_{-0.0091}$             \\
 $L_{\star}$        & [$L_{\odot}$]& $0.667^{+0.072}_{-0.061}$ \\
 $t_{\star}$        & [Gyr]     & $7.4^{+2.6}_{-2.4}$       \\
\hline
 \multicolumn{3}{c}{WASP-4 b} \\
\hline
 $T_0$              & [BJD]\tablefootmark{(a)}     & $4697.798269\pm0.000038$ \\
 $P$                & [days]    & $1.33823373\pm0.00000031$   \\
 $a$                & [AU]      & $0.02294^{+0.00041}_{-0.00040}$         \\
 $S$                & [$S_{\oplus}$]& $1268^{+100}_{-87}$    \\
 $T_{\mathrm{eq}}$\tablefootmark{(b)}  & [K]       & $1663^{+32}_{-29}$               \\
 $\mathrm{d}F$               &           & $0.023628^{+0.00011}_{-0.000098}$         \\
 $b$                & [$R_{\star}$] & $0.102^{+0.052}_{-0.065}$         \\
 $W$                & [min]     & $128.93^{+0.22}_{-0.20}$             \\
 $i_p$              & [$^{\circ}$] & $88.93^{+0.68}_{-0.56}$   \\
 $e$                &           & 0 (fixed)                 \\
 $\omega$\tablefootmark{(c)}           & [$^{\circ}$] & 0 (fixed)              \\
 $M_p$              & [$M_{\mathrm{J}}$] & $1.240^{+0.044}_{-0.043}$    \\
 $R_p$              & [$R_{\mathrm{J}}$] & $1.381\pm0.025$    \\
 $\rho_p$           & [g/cm$^3$] & $0.627\pm0.016$   \\        
\hline
\end{tabular}
\tablefoot{\tablefoottext{a}{TDB time standard, shifted by $-2'450'000$; }
\tablefoottext{b}{Assuming zero albedo; }
\tablefoottext{c}{Placed at the ascending node since the orbit is circular.}
}
\end{table}

\begin{table*}
\caption{WASP-4. Masses and radii of both the star and the planet derived by different authors.}             
\label{tab:Wasp4cmp}      
\centering          
\begin{tabular}{l l l l l l l }     
\hline\hline       
WASP-4 & \multicolumn{1}{c}{Our work} & \multicolumn{1}{c}{(1)} & \multicolumn{1}{c}{(2)} & \multicolumn{1}{c}{(3)} & \multicolumn{1}{c}{(4)} & \multicolumn{1}{c}{(5)} \\ 
\hline                    
 $R_{\star}$ [$R_{\odot}$] & $0.902\pm0.016$ & $0.907^{+0.014}_{-0.013}$ & $0.912\pm0.013$ & $0.914^{+0.024}_{-0.023}$ & $0.873^{+0.036}_{-0.027}$ & $0.937^{+0.040}_{-0.030}$ \\  
 $M_{\star}$ [$M_{\odot}$] & $0.898^{+0.048}_{-0.046}$ & $0.92\pm0.06$ & $0.925\pm0.040$ & $0.940^{+0.073}_{-0.069}$ & $0.85^{+0.11}_{-0.07}$ & $0.8997^{+0.077}_{-0.072}$ \\
 $R_p$ [$R_J$] & $1.381\pm0.025$ & $1.413\pm0.020$ & $1.365\pm0.021$ & $1.371^{+0.038}_{-0.035}$ & $1.304^{+0.054}_{-0.042}$ & $1.416^{+0.068}_{-0.043}$ \\
 $M_p$ [$M_J$] & $1.240^{+0.044}_{-0.043}$ &  & $1.237\pm0.064$ & $1.289^{+0.098}_{-0.073}$ & $1.21^{+0.13}_{-0.08}$ & $1.215^{+0.087}_{-0.079}$ \\
\hline                  
\end{tabular}
\tablefoot{(1): \citet{nikolov12}; (2): \citet{winn09}; (3): \citet{southworth09}; (4): \citet{gillon09a}; (5): \citet{wilson08}}
\end{table*}

\subsection{Comments on the uncertainties of isochronal parameters}\label{ssec:isoPrec}
Output parameters that are computed from isochrones (i.e. $t_{\star}$, $M_{\star}$ and $R_{\star}$) are given as if stellar evolutionary models are perfect. Therefore their respective errors are likely to be underestimated and should be considered as internal. One way to attribute more realistic uncertainties to isochronal parameters is by comparing our results with another set of evolutionary models. If the target stars and the interpolating algorithm are the same, differences arising in the output are to be attributed to the different input physics of the evolutionary models, such as the equations of state, adopted solar mixture, initial chemical composition of stars, nuclear reaction rates, opacities, overshooting treatment, mixing-length parameter, and atmospheric models. Setting up all these ingredients depends on our knowledge of stellar evolution, different choices produce different models, and consequent variations in the output parameter estimations are a measure of the uncertainty of evolutionary models.

Our isochrone placement as autonomous routine is also able to interact with the grids of tracks and isochrones produced by CLES \citep[Code Liégeois d'Évolution Stellaire;][]{scuflaire08}. These models span just a limited range in mass ($0.90 _{\odot}\le M \le 1.30 M_{\odot}$) and metallicity ($0.008\le Z \le 0.018$) so far: while HD 219134 is outside the mass range, luckily WASP-4 falls inside. Among the several differences in terms of stellar input physics between PARSEC and CLES models, one is related to the helium content $Y$. In PARSEC models, $Y$ is assumed to increase with $Z$, according to
\begin{equation}
 Y=Y_p+\frac{\Delta Y}{\Delta Z}Z
 \label{eq:Y}
,\end{equation}
where $Y_p=0.2485$ \citep{komatsu11} is the primordial helium abundance, while the helium-to-metal enrichment ratio $\frac{\Delta Y}{\Delta Z}=1.78$ is derived from solar calibration. Instead, $\frac{\Delta Y}{\Delta Z}$ is not fixed in CLES, such that $Y$ is not a function of $Z$, and actually for any given value of $Z$, four possible values of $Y$ are available, from $Y=0.25$ to $Y=0.28$ at steps of 0.01.

To evaluate the impact of stellar model systematics, we decided to analyse WASP-4 through our autonomous isochrone placement routine. Input parameters were $T_{\mathrm{eff}}$, [Fe/H], $\rho_{\star}$ as derived from transit, and $v\sin{i_{\star}}$. Interpolations were performed considering both PARSEC models and the four sets of CLES grids (one per each $Y$ value). The helium content cannot be generally constrained unless high quality asteroseismic oscillation frequencies are available \citep[see e.g.][]{lebreton14,buldgen16}.

According to eq. (\ref{eq:Y}), after considering element diffusion, PARSEC models assign an initial helium content $Y=0.28$ to WASP-4. By comparing the output $t_{\star}$, $M_{\star}$, and $R_{\star}$ values derived by the two models at the same $Y=0.28$ content, the relative differences (CLES versus PARSEC) in age, mass, and radius are $-19\%$, $+1.3\%$ and $+0.5\%$, respectively. If we also consider the effect of helium content on theoretical models, output variations on age, mass, and radius rise to $-23\%$, $+5.8\%$ and $+1.9\%$, respectively.

From this test case, we deduce that for having realistic error bars on the isochronal output parameters, conservative reference systematics to be added to the internal uncertainties are of the order of $\sim20\%$, $\sim6\%$, and $\sim2\%$ for age, mass, and radius, respectively. A complete study about isochrone precision is beyond the scope of the paper; properly comparing two different evolutionary models would require a wide stellar sample, possibly spanning different locations in the HRD, while the just mentioned percentage values are provided to give an initial idea about isochrone precision.

\section{Conclusions}\label{sec:conclusions}
We presented our MCMCI code that was born by merging the MCMC code \citep[see e.g.][]{gillon10,gillon12} and the isochrone placement algorithm developed by \citet{bonfanti15,bonfanti16}. As a reference, we took the opportunity to recall the working details of the two codes separately considering their most recent implementations.

Summing up, as the name suggests, the MCMC code is based upon Markov chain Monte Carlo statistics and it aims to retrieve the main parameters of an exoplanetary system once that photometric or RV time series are provided and stellar observational parameters are available. Given the indirect nature of the transit and RV methods, $M_{\star}$ and $R_{\star}$ have to  be  known to derive the mass and radius of the planet, which is possible as a consequence of stellar evolutionary models.
Starting from this necessity, we integrated isochrone placement into the MCMC, thereby building a new MCMCI code, which at each step of the chain ``asks'' theoretical models to provide $M_{\star}$ and $R_{\star}$, given the available observational data.

The main advantage is dealing with a powerful tool that retrieves PDFs of all the planetary and stellar parameters at a time without the need to split the analysis for separately recovering theoretical stellar parameters. 
In particular, the derived PDFs of $M_{\star}$ and $R_{\star}$ are a strict consequence of the values that are obtained at each step of the chains from the perturbation of the observational parameters.
In addition, our code is also able to establish the stellar age $t_{\star}$, considering the constraints coming both from stellar evolutionary models and several empirical age relations.

The MCMCI was tested by selecting two already studied planetary systems that may be considered representative of two different reference cases.

First we considered HD 219134, which hosts several planets (four according to the model we followed), two of which are transiting super Earths. This is the typical situation of shallow transits for which the transit impact parameter $b$ cannot be directly inferred from the LCs and knowledge of $R_{\star}$ is fundamental. We followed two different approaches, both considering an interferometric measurement of $R$ and also recovering it once $T_{\mathrm{eff}}$ was calibrated from $B-V$ and $L$ was deduced from $V$ magnitude and distance.

Second we considered WASP-4, that is known to host a hot Jupiter. The transit features are evident such that $\rho_{\star}$ can be inferred from the LCs and it enters the set of input parameters for inferring $M_{\star}$ and $R_{\star}$ from theoretical models.

Very good agreement was found by comparing our results to those reported in the literature, which suggests that our tool represents a powerful and integrated solution to analyse an exoplanetary system fully giving refined and reliable stellar and planetary parameters at the same time. In particular, planetary mass values are more precise if compared with previous determinations available in the literature.

Aside from the specific exoplanetary context, we also took the opportunity to speculate about isochrone precision in this specific work. By employing PARSEC and CLES stellar evolutionary models for studying WASP-4, we quantify the systematics affecting isochronal age ($\delta t_{\star}\sim20\%$), mass ($\delta M_{\star}\sim6\%$), and radius ($\delta R_{\star}\sim2\%$). Although we are aware that these percentage values should be considered as just indicative estimates of the isochrone precision because they are inferred from a single test case, these values draw attention to the fact that any isochronal parameter suffers intrinsic uncertainties, which are related to our current ability to model stars.

\begin{acknowledgements}
This work is based in part on observations made with the {\it Spitzer Space Telescope}, which is operated by the Jet Propulsion Laboratory, California Institute of Technology under a contract with NASA. M. Gillon is Senior Research Associate at F.R.S-FNRS. The research leading to these results has received funding from the ARC grant for Concerted Research Actions, financed by the Wallonia-Brussels Federation. A.B. is a postodoctoral researcher funded by the ESA PRODEX programme for CHEOPS (PEA 4000113509). The authors thank the referee, Jason Eastman; all his suggestions and comments lead to a remarkable improvement of the paper. Moreover, the authors thank A. Noëls, V. Van Grootel and A. Collier-Cameron for fruitful discussions that help improving the paper as well.

\end{acknowledgements}

%
%
\bibliography{biblio}

\begin{thebibliography}{124}
\expandafter\ifx\csname natexlab\endcsname\relax\def\natexlab#1{#1}\fi

\bibitem[{{Anderson} {et~al.}(2011){Anderson}, {Collier Cameron}, {Hellier},
  {Lendl}, {Maxted}, {Pollacco}, {Queloz}, {Smalley}, {Smith}, {Todd},
  {Triaud}, {West}, {Barros}, {Enoch}, {Gillon}, {Lister}, {Pepe},
  {S{\'e}gransan}, {Street}, \& {Udry}}]{anderson11}
{Anderson}, D.~R., {Collier Cameron}, A., {Hellier}, C., {et~al.} 2011, \apjl,
  726, L19

\bibitem[{{Angus} {et~al.}(2019){Angus}, {Morton}, {Foreman-Mackey}, {van
  Saders}, {Curtis}, {Kane}, {Bedell}, {Kiman}, {Hogg}, \& {Brewer}}]{angus19}
{Angus}, R., {Morton}, T.~D., {Foreman-Mackey}, D., {et~al.} 2019, \aj, 158,
  173

\bibitem[{{Auvergne} {et~al.}(2009){Auvergne}, {Bodin}, {Boisnard}, {Buey},
  {Chaintreuil}, {Epstein}, {Jouret}, {Lam-Trong}, {Levacher}, {Magnan},
  {Perez}, {Plasson}, {Plesseria}, {Peter}, {Steller}, {Tiph{\`e}ne}, {Baglin},
  {Agogu{\'e}}, {Appourchaux}, {Barbet}, {Beaufort}, {Bellenger}, {Berlin},
  {Bernardi}, {Blouin}, {Boumier}, {Bonneau}, {Briet}, {Butler}, {Cautain},
  {Chiavassa}, {Costes}, {Cuvilho}, {Cunha-Parro}, {de Oliveira Fialho},
  {Decaudin}, {Defise}, {Djalal}, {Docclo}, {Drummond}, {Dupuis}, {Exil},
  {Faur{\'e}}, {Gaboriaud}, {Gamet}, {Gavalda}, {Grolleau}, {Gueguen},
  {Guivarc'h}, {Guterman}, {Hasiba}, {Huntzinger}, {Hustaix}, {Imbert},
  {Jeanville}, {Johlander}, {Jorda}, {Journoud}, {Karioty}, {Kerjean},
  {Lafond}, {Lapeyrere}, {Landiech}, {Larqu{\'e}}, {Laudet}, {Le Merrer},
  {Leporati}, {Leruyet}, {Levieuge}, {Llebaria}, {Martin}, {Mazy}, {Mesnager},
  {Michel}, {Moalic}, {Monjoin}, {Naudet}, {Neukirchner}, {Nguyen-Kim},
  {Ollivier}, {Orcesi}, {Ottacher}, {Oulali}, {Parisot}, {Perruchot},
  {Piacentino}, {Pinheiro da Silva}, {Platzer}, {Pontet}, {Pradines},
  {Quentin}, {Rohbeck}, {Rolland}, {Rollenhagen}, {Romagnan}, {Russ}, {Samadi},
  {Schmidt}, {Schwartz}, {Sebbag}, {Smit}, {Sunter}, {Tello}, {Toulouse},
  {Ulmer}, {Vandermarcq}, {Vergnault}, {Wallner}, {Waultier}, \&
  {Zanatta}}]{auvergne09}
{Auvergne}, M., {Bodin}, P., {Boisnard}, L., {et~al.} 2009, \aap, 506, 411

\bibitem[{{Baluev}(2018)}]{baluev18}
{Baluev}, R.~V. 2018, Astronomy and Computing, 25, 221

\bibitem[{{Baranne} {et~al.}(1996){Baranne}, {Queloz}, {Mayor}, {Adrianzyk},
  {Knispel}, {Kohler}, {Lacroix}, {Meunier}, {Rimbaud}, \& {Vin}}]{baranne96}
{Baranne}, A., {Queloz}, D., {Mayor}, M., {et~al.} 1996, \aaps, 119, 373

\bibitem[{{Barnes}(2010)}]{barnes10}
{Barnes}, S.~A. 2010, \apj, 722, 222

\bibitem[{{Barrag{\'a}n} {et~al.}(2019){Barrag{\'a}n}, {Gandolfi}, \&
  {Antoniciello}}]{barragan19}
{Barrag{\'a}n}, O., {Gandolfi}, D., \& {Antoniciello}, G. 2019, \mnras, 482,
  1017

\bibitem[{{Basri} {et~al.}(2005){Basri}, {Borucki}, \& {Koch}}]{basri05}
{Basri}, G., {Borucki}, W.~J., \& {Koch}, D. 2005, \nar, 49, 478

\bibitem[{{Beerer} {et~al.}(2011){Beerer}, {Knutson}, {Burrows}, {Fortney},
  {Agol}, {Charbonneau}, {Cowan}, {Deming}, {Desert}, {Langton}, {Laughlin},
  {Lewis}, \& {Showman}}]{beerer11}
{Beerer}, I.~M., {Knutson}, H.~A., {Burrows}, A., {et~al.} 2011, \apj, 727, 23

\bibitem[{{Betancourt}(2017)}]{betancourt17}
{Betancourt}, M. 2017, arXiv e-prints [\eprint[arXiv]{1701.02434}]

\bibitem[{{Bonfanti} {et~al.}(2016){Bonfanti}, {Ortolani}, \&
  {Nascimbeni}}]{bonfanti16}
{Bonfanti}, A., {Ortolani}, S., \& {Nascimbeni}, V. 2016, \aap, 585, A5

\bibitem[{{Bonfanti} {et~al.}(2015){Bonfanti}, {Ortolani}, {Piotto}, \&
  {Nascimbeni}}]{bonfanti15}
{Bonfanti}, A., {Ortolani}, S., {Piotto}, G., \& {Nascimbeni}, V. 2015, \aap,
  575, A18

\bibitem[{{Borucki} {et~al.}(2010){Borucki}, {Koch}, {Basri}, {Batalha},
  {Brown}, {Caldwell}, {Caldwell}, {Christensen-Dalsgaard}, {Cochran},
  {DeVore}, {Dunham}, {Dupree}, {Gautier}, {Geary}, {Gilliland}, {Gould},
  {Howell}, {Jenkins}, {Kondo}, {Latham}, {Marcy}, {Meibom}, {Kjeldsen},
  {Lissauer}, {Monet}, {Morrison}, {Sasselov}, {Tarter}, {Boss}, {Brownlee},
  {Owen}, {Buzasi}, {Charbonneau}, {Doyle}, {Fortney}, {Ford}, {Holman},
  {Seager}, {Steffen}, {Welsh}, {Rowe}, {Anderson}, {Buchhave}, {Ciardi},
  {Walkowicz}, {Sherry}, {Horch}, {Isaacson}, {Everett}, {Fischer}, {Torres},
  {Johnson}, {Endl}, {MacQueen}, {Bryson}, {Dotson}, {Haas}, {Kolodziejczak},
  {Van Cleve}, {Chandrasekaran}, {Twicken}, {Quintana}, {Clarke}, {Allen},
  {Li}, {Wu}, {Tenenbaum}, {Verner}, {Bruhweiler}, {Barnes}, \&
  {Prsa}}]{borucki10}
{Borucki}, W.~J., {Koch}, D., {Basri}, G., {et~al.} 2010, Science, 327, 977

\bibitem[{{Bouma} {et~al.}(2019){Bouma}, {Winn}, {Baxter}, {Bhatti}, {Dai},
  {Daylan}, {D{\'e}sert}, {Hill}, {Kane}, {Stassun}, {Villasenor}, {Ricker},
  {Vanderspek}, {Latham}, {Seager}, {Jenkins}, {Berta-Thompson}, {Col{\'o}n},
  {Fausnaugh}, {Glidden}, {Guerrero}, {Rodriguez}, {Twicken}, \&
  {Wohler}}]{bouma19}
{Bouma}, L.~G., {Winn}, J.~N., {Baxter}, C., {et~al.} 2019, \aj, 157, 217

\bibitem[{{Boyajian} {et~al.}(2012){Boyajian}, {von Braun}, {van Belle},
  {McAlister}, {ten Brummelaar}, {Kane}, {Muirhead}, {Jones}, {White},
  {Schaefer}, {Ciardi}, {Henry}, {L{\'o}pez-Morales}, {Ridgway}, {Gies}, {Jao},
  {Rojas-Ayala}, {Parks}, {Sturmann}, {Sturmann}, {Turner}, {Farrington},
  {Goldfinger}, \& {Berger}}]{boyajian12}
{Boyajian}, T.~S., {von Braun}, K., {van Belle}, G., {et~al.} 2012, \apj, 757,
  112

\bibitem[{{Bressan} {et~al.}(2012){Bressan}, {Marigo}, {Girardi}, {Salasnich},
  {Dal Cero}, {Rubele}, \& {Nanni}}]{bressan12}
{Bressan}, A., {Marigo}, P., {Girardi}, L., {et~al.} 2012, \mnras, 427, 127

\bibitem[{{Broeg} {et~al.}(2013){Broeg}, {Fortier}, {Ehrenreich}, {Alibert},
  {Baumjohann}, {Benz}, {Deleuil}, {Gillon}, {Ivanov}, {Liseau}, {Meyer},
  {Oloffson}, {Pagano}, {Piotto}, {Pollacco}, {Queloz}, {Ragazzoni}, {Renotte},
  {Steller}, \& {Thomas}}]{broeg13}
{Broeg}, C., {Fortier}, A., {Ehrenreich}, D., {et~al.} 2013, in European
  Physical Journal Web of Conferences, Vol.~47, European Physical Journal Web
  of Conferences, 03005

\bibitem[{{Buldgen} {et~al.}(2016){Buldgen}, {Reese}, \& {Dupret}}]{buldgen16}
{Buldgen}, G., {Reese}, D.~R., \& {Dupret}, M.~A. 2016, \aap, 585, A109

\bibitem[{{Burgers}(1969)}]{burgers69}
{Burgers}, J. 1969, {Flow equations for composites gases} (New York: Academic
  Press)

\bibitem[{{Caffau} {et~al.}(2011){Caffau}, {Ludwig}, {Steffen}, {Freytag}, \&
  {Bonifacio}}]{caffau11}
{Caffau}, E., {Ludwig}, H.-G., {Steffen}, M., {Freytag}, B., \& {Bonifacio}, P.
  2011, \solphys, 268, 255

\bibitem[{Casella \& George(1992)}]{casella92}
Casella, G. \& George, E.~I. 1992, The American Statistician, 46, 167

\bibitem[{{Chaboyer} {et~al.}(2001){Chaboyer}, {Fenton}, {Nelan}, {Patnaude},
  \& {Simon}}]{chaboyer01}
{Chaboyer}, B., {Fenton}, W.~H., {Nelan}, J.~E., {Patnaude}, D.~J., \& {Simon},
  F.~E. 2001, \apj, 562, 521

\bibitem[{{Chaplin} {et~al.}(2014){Chaplin}, {Basu}, {Huber}, {Serenelli},
  {Casagrande}, {Silva Aguirre}, {Ball}, {Creevey}, {Gizon}, {Handberg},
  {Karoff}, {Lutz}, {Marques}, {Miglio}, {Stello}, {Suran}, {Pricopi},
  {Metcalfe}, {Monteiro}, {Molenda-{\.Z}akowicz}, {Appourchaux},
  {Christensen-Dalsgaard}, {Elsworth}, {Garc{\'{\i}}a}, {Houdek}, {Kjeldsen},
  {Bonanno}, {Campante}, {Corsaro}, {Gaulme}, {Hekker}, {Mathur}, {Mosser},
  {R{\'e}gulo}, \& {Salabert}}]{chaplin14}
{Chaplin}, W.~J., {Basu}, S., {Huber}, D., {et~al.} 2014, \apjs, 210, 1

\bibitem[{{Chapman} \& {Cowling}(1970, 3rd~ed.)}]{chapman70}
{Chapman}, S. \& {Cowling}, T.~G. 1970, 3rd~ed., {The mathematical theory of
  non-uniform gases. an account of the kinetic theory of viscosity, thermal
  conduction and diffusion in gases} (Cambridge: University Press)

\bibitem[{{Charbonneau} {et~al.}(2008){Charbonneau}, {Knutson}, {Barman},
  {Allen}, {Mayor}, {Megeath}, {Queloz}, \& {Udry}}]{charbonneau08}
{Charbonneau}, D., {Knutson}, H.~A., {Barman}, T., {et~al.} 2008, \apj, 686,
  1341

\bibitem[{{Chen} {et~al.}(2014){Chen}, {Girardi}, {Bressan}, {Marigo},
  {Barbieri}, \& {Kong}}]{chen14}
{Chen}, Y., {Girardi}, L., {Bressan}, A., {et~al.} 2014, \mnras, 444, 2525

\bibitem[{{Claret} \& {Bloemen}(2011)}]{claret11}
{Claret}, A. \& {Bloemen}, S. 2011, \aap, 529, A75

\bibitem[{{Collier Cameron} {et~al.}(2010){Collier Cameron}, {Bruce}, {Miller},
  {Triaud}, \& {Queloz}}]{collierCameron10}
{Collier Cameron}, A., {Bruce}, V.~A., {Miller}, G.~R.~M., {Triaud},
  A.~H.~M.~J., \& {Queloz}, D. 2010, \mnras, 403, 151

\bibitem[{{Cosentino} {et~al.}(2012){Cosentino}, {Lovis}, {Pepe}, {Collier
  Cameron}, {Latham}, {Molinari}, {Udry}, {Bezawada}, {Black}, {Born},
  {Buchschacher}, {Charbonneau}, {Figueira}, {Fleury}, {Galli}, {Gallie},
  {Gao}, {Ghedina}, {Gonzalez}, {Gonzalez}, {Guerra}, {Henry}, {Horne},
  {Hughes}, {Kelly}, {Lodi}, {Lunney}, {Maire}, {Mayor}, {Micela}, {Ordway},
  {Peacock}, {Phillips}, {Piotto}, {Pollacco}, {Queloz}, {Rice}, {Riverol},
  {Riverol}, {San Juan}, {Sasselov}, {Segransan}, {Sozzetti}, {Sosnowska},
  {Stobie}, {Szentgyorgyi}, {Vick}, \& {Weber}}]{cosentino12}
{Cosentino}, R., {Lovis}, C., {Pepe}, F., {et~al.} 2012, in \procspie, Vol.
  8446, Ground-based and Airborne Instrumentation for Astronomy IV, 84461V

\bibitem[{{da Silva} {et~al.}(2012){da Silva}, {Porto de Mello}, {Milone}, {da
  Silva}, {Ribeiro}, \& {Rocha-Pinto}}]{daSilva12}
{da Silva}, R., {Porto de Mello}, G.~F., {Milone}, A.~C., {et~al.} 2012, \aap,
  542, A84

\bibitem[{{Davenport} {et~al.}(2014){Davenport}, {Hawley}, {Hebb},
  {Wisniewski}, {Kowalski}, {Johnson}, {Malatesta}, {Peraza}, {Keil},
  {Silverberg}, {Jansen}, {Scheffler}, {Berdis}, {Larsen}, \&
  {Hilton}}]{davenport14}
{Davenport}, J. R.~A., {Hawley}, S.~L., {Hebb}, L., {et~al.} 2014, \apj, 797,
  122

\bibitem[{{Dawson} \& {Johnson}(2012)}]{dawson12}
{Dawson}, R.~I. \& {Johnson}, J.~A. 2012, \apj, 756, 122

\bibitem[{{Dawson} \& {Johnson}(2018)}]{dawson18}
{Dawson}, R.~I. \& {Johnson}, J.~A. 2018, \araa, 56, 175

\bibitem[{{Denissenkov}(2010)}]{denissenkov10}
{Denissenkov}, P.~A. 2010, \apj, 719, 28

\bibitem[{{Dotter}(2016)}]{dotter16}
{Dotter}, A. 2016, \apjs, 222, 8

\bibitem[{{Dotter} {et~al.}(2017){Dotter}, {Conroy}, {Cargile}, \&
  {Asplund}}]{dotter17}
{Dotter}, A., {Conroy}, C., {Cargile}, P., \& {Asplund}, M. 2017, \apj, 840, 99

\bibitem[{{Eastman} {et~al.}(2013){Eastman}, {Gaudi}, \& {Agol}}]{eastman13}
{Eastman}, J., {Gaudi}, B.~S., \& {Agol}, E. 2013, \pasp, 125, 83

\bibitem[{Eastman {et~al.}(2010)Eastman, Siverd, \& Scott~Gaudi}]{eastman10}
Eastman, J., Siverd, R., \& Scott~Gaudi, B. 2010, Publications of the
  Astronomical Society of the Pacific, 122, 935

\bibitem[{{Eastman} {et~al.}(2019){Eastman}, {Rodriguez}, {Agol}, {Stassun},
  {Beatty}, {Vanderburg}, {Gaudi}, {Collins}, \& {Luger}}]{eastman19}
{Eastman}, J.~D., {Rodriguez}, J.~E., {Agol}, E., {et~al.} 2019, arXiv e-prints
  [\eprint[arXiv]{1907.09480}]

\bibitem[{{Espinoza} {et~al.}(2018){Espinoza}, {Kossakowski}, \&
  {Brahm}}]{espinoza18}
{Espinoza}, N., {Kossakowski}, D., \& {Brahm}, R. 2018, arXiv e-prints
  [\eprint[arXiv]{1812.08549}]

\bibitem[{{Fazio} {et~al.}(2004){Fazio}, {Hora}, {Allen}, {Ashby}, {Barmby},
  {Deutsch}, {Huang}, {Kleiner}, {Marengo}, {Megeath}, {Melnick}, {Pahre},
  {Patten}, {Polizotti}, {Smith}, {Taylor}, {Wang}, {Willner}, {Hoffmann},
  {Pipher}, {Forrest}, {McMurty}, {McCreight}, {McKelvey}, {McMurray}, {Koch},
  {Moseley}, {Arendt}, {Mentzell}, {Marx}, {Losch}, {Mayman}, {Eichhorn},
  {Krebs}, {Jhabvala}, {Gezari}, {Fixsen}, {Flores}, {Shakoorzadeh}, {Jungo},
  {Hakun}, {Workman}, {Karpati}, {Kichak}, {Whitley}, {Mann}, {Tollestrup},
  {Eisenhardt}, {Stern}, {Gorjian}, {Bhattacharya}, {Carey}, {Nelson},
  {Glaccum}, {Lacy}, {Lowrance}, {Laine}, {Reach}, {Stauffer}, {Surace},
  {Wilson}, {Wright}, {Hoffman}, {Domingo}, \& {Cohen}}]{fazio04}
{Fazio}, G.~G., {Hora}, J.~L., {Allen}, L.~E., {et~al.} 2004, \apjs, 154, 10

\bibitem[{{Feltzing} {et~al.}(2017){Feltzing}, {Howes}, {McMillan}, \&
  {Stonkut{\.e}}}]{feltzing17}
{Feltzing}, S., {Howes}, L.~M., {McMillan}, P.~J., \& {Stonkut{\.e}}, E. 2017,
  \mnras, 465, L109

\bibitem[{{Ford}(2005)}]{ford05}
{Ford}, E.~B. 2005, \aj, 129, 1706

\bibitem[{{Ford}(2006)}]{ford06}
{Ford}, E.~B. 2006, \apj, 642, 505

\bibitem[{{Foreman-Mackey} {et~al.}(2017){Foreman-Mackey}, {Agol},
  {Ambikasaran}, \& {Angus}}]{foreman17}
{Foreman-Mackey}, D., {Agol}, E., {Ambikasaran}, S., \& {Angus}, R. 2017, \aj,
  154, 220

\bibitem[{{Foreman-Mackey} {et~al.}(2013){Foreman-Mackey}, {Hogg}, {Lang}, \&
  {Goodman}}]{foreman13}
{Foreman-Mackey}, D., {Hogg}, D.~W., {Lang}, D., \& {Goodman}, J. 2013,
  Publications of the Astronomical Society of the Pacific, 125, 306

\bibitem[{{Gaia Collab.} {et~al.}(2018){Gaia Collab.}, {Brown}, {Vallenari},
  {Prusti}, {de Bruijne}, {Babusiaux}, {Bailer-Jones}, {Biermann}, {Evans},
  {Eyer}, {Jansen}, {Jordi}, {Klioner}, {Lammers}, {Lindegren}, {Luri},
  {Mignard}, {Panem}, {Pourbaix}, {Randich}, {Sartoretti}, {Siddiqui},
  {Soubiran}, {van Leeuwen}, {Walton}, {Arenou}, {Bastian}, {Cropper},
  {Drimmel}, {Katz}, {Lattanzi}, {Bakker}, {Cacciari}, {Casta{\~n}eda},
  {Chaoul}, {Cheek}, {De Angeli}, {Fabricius}, {Guerra}, {Holl}, {Masana},
  {Messineo}, {Mowlavi}, {Nienartowicz}, {Panuzzo}, {Portell}, {Riello},
  {Seabroke}, {Tanga}, {Th{\'e}venin}, {Gracia-Abril}, {Comoretto},
  {Garcia-Reinaldos}, {Teyssier}, {Altmann}, {Andrae}, {Audard},
  {Bellas-Velidis}, {Benson}, {Berthier}, {Blomme}, {Burgess}, {Busso},
  {Carry}, {Cellino}, {Clementini}, {Clotet}, {Creevey}, {Davidson}, {De
  Ridder}, {Delchambre}, {Dell'Oro}, {Ducourant},
  {Fern{\'a}ndez-Hern{\'a}ndez}, {Fouesneau}, {Fr{\'e}mat}, {Galluccio},
  {Garc{\'\i}a-Torres}, {Gonz{\'a}lez-N{\'u}{\~n}ez}, {Gonz{\'a}lez-Vidal},
  {Gosset}, {Guy}, {Halbwachs}, {Hambly}, {Harrison}, {Hern{\'a}ndez},
  {Hestroffer}, {Hodgkin}, {Hutton}, {Jasniewicz}, {Jean-Antoine-Piccolo},
  {Jordan}, {Korn}, {Krone-Martins}, {Lanzafame}, {Lebzelter}, {L{\"o}ffler},
  {Manteiga}, {Marrese}, {Mart{\'\i}n-Fleitas}, {Moitinho}, {Mora}, {Muinonen},
  {Osinde}, {Pancino}, {Pauwels}, {Petit}, {Recio-Blanco}, {Richards},
  {Rimoldini}, {Robin}, {Sarro}, {Siopis}, {Smith}, {Sozzetti}, {S{\"u}veges},
  {Torra}, {van Reeven}, {Abbas}, {Abreu Aramburu}, {Accart}, {Aerts},
  {Altavilla}, {{\'A}lvarez}, {Alvarez}, {Alves}, {Anderson}, {Andrei},
  {Anglada Varela}, {Antiche}, {Antoja}, {Arcay}, {Astraatmadja}, {Bach},
  {Baker}, {Balaguer-N{\'u}{\~n}ez}, {Balm}, {Barache}, {Barata}, {Barbato},
  {Barblan}, {Barklem}, {Barrado}, {Barros}, {Barstow}, {Bartholom{\'e}
  Mu{\~n}oz}, {Bassilana}, {Becciani}, {Bellazzini}, {Berihuete}, {Bertone},
  {Bianchi}, {Bienaym{\'e}}, {Blanco-Cuaresma}, {Boch}, {Boeche}, {Bombrun},
  {Borrachero}, {Bossini}, {Bouquillon}, {Bourda}, {Bragaglia}, {Bramante},
  {Breddels}, {Bressan}, {Brouillet}, {Br{\"u}semeister}, {Brugaletta},
  {Bucciarelli}, {Burlacu}, {Busonero}, {Butkevich}, {Buzzi}, {Caffau},
  {Cancelliere}, {Cannizzaro}, {Cantat-Gaudin}, {Carballo}, {Carlucci},
  {Carrasco}, {Casamiquela}, {Castellani}, {Castro-Ginard}, {Charlot},
  {Chemin}, {Chiavassa}, {Cocozza}, {Costigan}, {Cowell}, {Crifo}, {Crosta},
  {Crowley}, {Cuypers}, {Dafonte}, {Damerdji}, {Dapergolas}, {David}, {David},
  {de Laverny}, {De Luise}, {De March}, {de Martino}, {de Souza}, {de Torres},
  {Debosscher}, {del Pozo}, {Delbo}, {Delgado}, {Delgado}, {Di Matteo},
  {Diakite}, {Diener}, {Distefano}, {Dolding}, {Drazinos}, {Dur{\'a}n},
  {Edvardsson}, {Enke}, {Eriksson}, {Esquej}, {Eynard Bontemps}, {Fabre},
  {Fabrizio}, {Faigler}, {Falc{\~a}o}, {Farr{\`a}s Casas}, {Federici},
  {Fedorets}, {Fernique}, {Figueras}, {Filippi}, {Findeisen}, {Fonti},
  {Fraile}, {Fraser}, {Fr{\'e}zouls}, {Gai}, {Galleti}, {Garabato},
  {Garc{\'\i}a-Sedano}, {Garofalo}, {Garralda}, {Gavel}, {Gavras}, {Gerssen},
  {Geyer}, {Giacobbe}, {Gilmore}, {Girona}, {Giuffrida}, {Glass}, {Gomes},
  {Granvik}, {Gueguen}, {Guerrier}, {Guiraud}, {Guti{\'e}rrez-S{\'a}nchez},
  {Haigron}, {Hatzidimitriou}, {Hauser}, {Haywood}, {Heiter}, {Helmi}, {Heu},
  {Hilger}, {Hobbs}, {Hofmann}, {Holland}, {Huckle}, {Hypki}, {Icardi},
  {Jan{\ss}en}, {Jevardat de Fombelle}, {Jonker}, {Juh{\'a}sz}, {Julbe},
  {Karampelas}, {Kewley}, {Klar}, {Kochoska}, {Kohley}, {Kolenberg},
  {Kontizas}, {Kontizas}, {Koposov}, {Kordopatis}, {Kostrzewa-Rutkowska},
  {Koubsky}, {Lambert}, {Lanza}, {Lasne}, {Lavigne}, {Le Fustec}, {Le
  Poncin-Lafitte}, {Lebreton}, {Leccia}, {Leclerc}, {Lecoeur-Taibi},
  {Lenhardt}, {Leroux}, {Liao}, {Licata}, {Lindstr{\o}m}, {Lister}, {Livanou},
  {Lobel}, {L{\'o}pez}, {Managau}, {Mann}, {Mantelet}, {Marchal}, {Marchant},
  {Marconi}, {Marinoni}, {Marschalk{\'o}}, {Marshall}, {Martino}, {Marton},
  {Mary}, {Massari}, {Matijevi{\v{c}}}, {Mazeh}, {McMillan}, {Messina},
  {Michalik}, {Millar}, {Molina}, {Molinaro}, {Moln{\'a}r}, {Montegriffo},
  {Mor}, {Morbidelli}, {Morel}, {Morris}, {Mulone}, {Muraveva}, {Musella},
  {Nelemans}, {Nicastro}, {Noval}, {O'Mullane}, {Ord{\'e}novic},
  {Ord{\'o}{\~n}ez-Blanco}, {Osborne}, {Pagani}, {Pagano}, {Pailler},
  {Palacin}, {Palaversa}, {Panahi}, {Pawlak}, {Piersimoni}, {Pineau}, {Plachy},
  {Plum}, {Poggio}, {Poujoulet}, {Pr{\v{s}}a}, {Pulone}, {Racero}, {Ragaini},
  {Rambaux}, {Ramos-Lerate}, {Regibo}, {Reyl{\'e}}, {Riclet}, {Ripepi}, {Riva},
  {Rivard}, {Rixon}, {Roegiers}, {Roelens}, {Romero-G{\'o}mez}, {Rowell},
  {Royer}, {Ruiz-Dern}, {Sadowski}, {Sagrist{\`a} Sell{\'e}s}, {Sahlmann},
  {Salgado}, {Salguero}, {Sanna}, {Santana-Ros}, {Sarasso}, {Savietto},
  {Schultheis}, {Sciacca}, {Segol}, {Segovia}, {S{\'e}gransan}, {Shih},
  {Siltala}, {Silva}, {Smart}, {Smith}, {Solano}, {Solitro}, {Sordo}, {Soria
  Nieto}, {Souchay}, {Spagna}, {Spoto}, {Stampa}, {Steele},
  {Steidelm{\"u}ller}, {Stephenson}, {Stoev}, {Suess}, {Surdej}, {Szabados},
  {Szegedi-Elek}, {Tapiador}, {Taris}, {Tauran}, {Taylor}, {Teixeira},
  {Terrett}, {Teyssand ier}, {Thuillot}, {Titarenko}, {Torra Clotet}, {Turon},
  {Ulla}, {Utrilla}, {Uzzi}, {Vaillant}, {Valentini}, {Valette}, {van Elteren},
  {Van Hemelryck}, {van Leeuwen}, {Vaschetto}, {Vecchiato}, {Veljanoski},
  {Viala}, {Vicente}, {Vogt}, {von Essen}, {Voss}, {Votruba}, {Voutsinas},
  {Walmsley}, {Weiler}, {Wertz}, {Wevers}, {Wyrzykowski}, {Yoldas},
  {{\v{Z}}erjal}, {Ziaeepour}, {Zorec}, {Zschocke}, {Zucker}, {Zurbach}, \&
  {Zwitter}}]{GaiaDR2_18}
{Gaia Collab.}, {Brown}, A.~G.~A., {Vallenari}, A., {et~al.} 2018, \aap, 616,
  A1

\bibitem[{{Gelman} \& {Rubin}(1992)}]{gelman92}
{Gelman}, A. \& {Rubin}, D.~B. 1992, Statistical Science, 7, 457

\bibitem[{{Gillon} {et~al.}(2009{\natexlab{a}}){Gillon}, {Demory}, {Triaud},
  {Barman}, {Hebb}, {Montalb{\'a}n}, {Maxted}, {Queloz}, {Deleuil}, \&
  {Magain}}]{gillon09b}
{Gillon}, M., {Demory}, B.-O., {Triaud}, A.~H.~M.~J., {et~al.}
  2009{\natexlab{a}}, \aap, 506, 359

\bibitem[{{Gillon} {et~al.}(2017){Gillon}, {Demory}, {Van Grootel}, {Motalebi},
  {Lovis}, {Cameron}, {Charbonneau}, {Latham}, {Molinari}, {Pepe},
  {S{\'e}gransan}, {Sasselov}, {Udry}, {Mayor}, {Micela}, {Piotto}, \&
  {Sozzetti}}]{gillon17}
{Gillon}, M., {Demory}, B.-O., {Van Grootel}, V., {et~al.} 2017, Nature
  Astronomy, 1, 0056

\bibitem[{{Gillon} {et~al.}(2011){Gillon}, {Doyle}, {Lendl}, {Maxted},
  {Triaud}, {Anderson}, {Barros}, {Bento}, {Collier-Cameron}, {Enoch}, {Faedi},
  {Hellier}, {Jehin}, {Magain}, {Montalb{\'a}n}, {Pepe}, {Pollacco}, {Queloz},
  {Smalley}, {Segransan}, {Smith}, {Southworth}, {Udry}, {West}, \&
  {Wheatley}}]{gillon11}
{Gillon}, M., {Doyle}, A.~P., {Lendl}, M., {et~al.} 2011, \aap, 533, A88

\bibitem[{{Gillon} {et~al.}(2010){Gillon}, {Lanotte}, {Barman}, {Miller},
  {Demory}, {Deleuil}, {Montalb{\'a}n}, {Bouchy}, {Collier Cameron}, {Deeg},
  {Fortney}, {Fridlund}, {Harrington}, {Magain}, {Moutou}, {Queloz}, {Rauer},
  {Rouan}, \& {Schneider}}]{gillon10}
{Gillon}, M., {Lanotte}, A.~A., {Barman}, T., {et~al.} 2010, \aap, 511, A3

\bibitem[{{Gillon} {et~al.}(2009{\natexlab{b}}){Gillon}, {Smalley}, {Hebb},
  {Anderson}, {Triaud}, {Hellier}, {Maxted}, {Queloz}, \& {Wilson}}]{gillon09a}
{Gillon}, M., {Smalley}, B., {Hebb}, L., {et~al.} 2009{\natexlab{b}}, \aap,
  496, 259

\bibitem[{{Gillon} {et~al.}(2012){Gillon}, {Triaud}, {Fortney}, {Demory},
  {Jehin}, {Lendl}, {Magain}, {Kabath}, {Queloz}, {Alonso}, {Anderson},
  {Collier Cameron}, {Fumel}, {Hebb}, {Hellier}, {Lanotte}, {Maxted},
  {Mowlavi}, \& {Smalley}}]{gillon12}
{Gillon}, M., {Triaud}, A.~H.~M.~J., {Fortney}, J.~J., {et~al.} 2012, \aap,
  542, A4

\bibitem[{{Gim{\'e}nez}(2006)}]{gimenez06}
{Gim{\'e}nez}, A. 2006, \apj, 650, 408

\bibitem[{{G{\"u}nther} \& {Daylan}(2019)}]{guenther19}
{G{\"u}nther}, M.~N. \& {Daylan}, T. 2019, {allesfitter: Flexible star and
  exoplanet inference from photometry and radial velocity}

\bibitem[{{Hartman} {et~al.}(2019){Hartman}, {Bakos}, {Bayliss}, {Bento},
  {Bhatti}, {Brahm}, {Csubry}, {Espinoza}, {Henning}, {Jord{\'a}n}, {Mancini},
  {Penev}, {Rabus}, {Sarkis}, {Suc}, {de Val-Borro}, {Zhou}, {Addison},
  {Arriagada}, {Butler}, {Crane}, {Durkan}, {Shectman}, {Tan}, {Thompson},
  {Tinney}, {Wright}, {L{\'a}z{\'a}r}, {Papp}, \& {S{\'a}ri}}]{hartman19}
{Hartman}, J.~D., {Bakos}, G.~{\'A}., {Bayliss}, D., {et~al.} 2019, \aj, 157,
  55

\bibitem[{Hastings(1970)}]{hastings70}
Hastings, W.~K. 1970, Biometrika, 57, 97

\bibitem[{{Holman} \& {Murray}(2005)}]{holman05}
{Holman}, M.~J. \& {Murray}, N.~W. 2005, Science, 307, 1288

\bibitem[{{Holman} {et~al.}(2006){Holman}, {Winn}, {Latham}, {O'Donovan},
  {Charbonneau}, {Bakos}, {Esquerdo}, {Hergenrother}, {Everett}, \&
  {P{\'a}l}}]{holman06}
{Holman}, M.~J., {Winn}, J.~N., {Latham}, D.~W., {et~al.} 2006, \apj, 652, 1715

\bibitem[{{Howell} {et~al.}(2014){Howell}, {Sobeck}, {Haas}, {Still},
  {Barclay}, {Mullally}, {Troeltzsch}, {Aigrain}, {Bryson}, {Caldwell},
  {Chaplin}, {Cochran}, {Huber}, {Marcy}, {Miglio}, {Najita}, {Smith},
  {Twicken}, \& {Fortney}}]{howell14}
{Howell}, S.~B., {Sobeck}, C., {Haas}, M., {et~al.} 2014, \pasp, 126, 398

\bibitem[{Iglesias-Marzoa {et~al.}(2015)Iglesias-Marzoa, L{\'{o}}pez-Morales,
  \& Morales}]{iglesiasMarzoa15}
Iglesias-Marzoa, R., L{\'{o}}pez-Morales, M., \& Morales, M. J.~A. 2015,
  Publications of the Astronomical Society of the Pacific, 127, 567

\bibitem[{{J{\o}rgensen} \& {Lindegren}(2005)}]{jorgensen05}
{J{\o}rgensen}, B.~R. \& {Lindegren}, L. 2005, \aap, 436, 127

\bibitem[{{Knutson} {et~al.}(2008){Knutson}, {Charbonneau}, {Allen}, {Burrows},
  \& {Megeath}}]{knutson08}
{Knutson}, H.~A., {Charbonneau}, D., {Allen}, L.~E., {Burrows}, A., \&
  {Megeath}, S.~T. 2008, \apj, 673, 526

\bibitem[{{Komatsu} {et~al.}(2011){Komatsu}, {Smith}, {Dunkley}, {Bennett},
  {Gold}, {Hinshaw}, {Jarosik}, {Larson}, {Nolta}, {Page}, {Spergel},
  {Halpern}, {Hill}, {Kogut}, {Limon}, {Meyer}, {Odegard}, {Tucker}, {Weiland},
  {Wollack}, \& {Wright}}]{komatsu11}
{Komatsu}, E., {Smith}, K.~M., {Dunkley}, J., {et~al.} 2011, \apjs, 192, 18

\bibitem[{{Kopal}(1959)}]{kopal59}
{Kopal}, Z. 1959, {Close binary systems}

\bibitem[{{Kreidberg}(2015)}]{kreidberg15}
{Kreidberg}, L. 2015, \pasp, 127, 1161

\bibitem[{{Lebreton} \& {Goupil}(2014)}]{lebreton14}
{Lebreton}, Y. \& {Goupil}, M.~J. 2014, \aap, 569, A21

\bibitem[{{Loeb} \& {Gaudi}(2003)}]{loeb03}
{Loeb}, A. \& {Gaudi}, B.~S. 2003, \apjl, 588, L117

\bibitem[{M.~Neal(2012)}]{neal12}
M.~Neal, R. 2012, Handbook of Markov Chain Monte Carlo

\bibitem[{{Madhusudhan} \& {Winn}(2009)}]{madhusudhan09}
{Madhusudhan}, N. \& {Winn}, J.~N. 2009, \apj, 693, 784

\bibitem[{{Mamajek} \& {Hillenbrand}(2008)}]{mamajek08}
{Mamajek}, E.~E. \& {Hillenbrand}, L.~A. 2008, \apj, 687, 1264

\bibitem[{{Mandel} \& {Agol}(2002)}]{mandel02}
{Mandel}, K. \& {Agol}, E. 2002, \apjl, 580, L171

\bibitem[{{Marigo} {et~al.}(2013){Marigo}, {Bressan}, {Nanni}, {Girardi}, \&
  {Pumo}}]{marigo13}
{Marigo}, P., {Bressan}, A., {Nanni}, A., {Girardi}, L., \& {Pumo}, M.~L. 2013,
  \mnras, 434, 488

\bibitem[{{Marigo} {et~al.}(2017){Marigo}, {Girardi}, {Bressan}, {Rosenfield},
  {Aringer}, {Chen}, {Dussin}, {Nanni}, {Pastorelli}, {Rodrigues}, {Trabucchi},
  {Bladh}, {Dalcanton}, {Groenewegen}, {Montalb{\'a}n}, \& {Wood}}]{marigo17}
{Marigo}, P., {Girardi}, L., {Bressan}, A., {et~al.} 2017, \apj, 835, 77

\bibitem[{{Mayor} \& {Queloz}(1995)}]{mayor95}
{Mayor}, M. \& {Queloz}, D. 1995, \nat, 378, 355

\bibitem[{{Meibom} {et~al.}(2015){Meibom}, {Barnes}, {Platais}, {Gilliland},
  {Latham}, \& {Mathieu}}]{meibom15}
{Meibom}, S., {Barnes}, S.~A., {Platais}, I., {et~al.} 2015, \nat, 517, 589

\bibitem[{{Mishenina} {et~al.}(2008){Mishenina}, {Soubiran}, {Bienaym{\'e}},
  {Korotin}, {Belik}, {Usenko}, \& {Kovtyukh}}]{mishenina08}
{Mishenina}, T.~V., {Soubiran}, C., {Bienaym{\'e}}, O., {et~al.} 2008, \aap,
  489, 923

\bibitem[{{Morris}(1985)}]{morris85}
{Morris}, S.~L. 1985, \apj, 295, 143

\bibitem[{{Motalebi} {et~al.}(2015){Motalebi}, {Udry}, {Gillon}, {Lovis},
  {S{\'e}gransan}, {Buchhave}, {Demory}, {Malavolta}, {Dressing}, {Sasselov},
  {Rice}, {Charbonneau}, {Collier Cameron}, {Latham}, {Molinari}, {Pepe},
  {Affer}, {Bonomo}, {Cosentino}, {Dumusque}, {Figueira}, {Fiorenzano},
  {Gettel}, {Harutyunyan}, {Haywood}, {Johnson}, {Lopez}, {Lopez-Morales},
  {Mayor}, {Micela}, {Mortier}, {Nascimbeni}, {Philips}, {Piotto}, {Pollacco},
  {Queloz}, {Sozzetti}, {Vanderburg}, \& {Watson}}]{motalebi15}
{Motalebi}, F., {Udry}, S., {Gillon}, M., {et~al.} 2015, \aap, 584, A72

\bibitem[{{Murray} \& {Correia}(2011)}]{murray11}
{Murray}, C. \& {Correia}, A. 2011, Exoplanets, edited by S. Seager. Tucson,
  AZ: University of Arizona Press, 2011, 526 pp., ISBN 978-0-8165-2945-2., 15

\bibitem[{{Murray} \& {Dermott}(1999)}]{murray99}
{Murray}, C.~D. \& {Dermott}, S.~F. 1999, {Solar system dynamics}

\bibitem[{{Nikolov} {et~al.}(2012){Nikolov}, {Henning}, {Koppenhoefer},
  {Lendl}, {Maciejewski}, \& {Greiner}}]{nikolov12}
{Nikolov}, N., {Henning}, T., {Koppenhoefer}, J., {et~al.} 2012, \aap, 539,
  A159

\bibitem[{{Nissen}(2016)}]{nissen16}
{Nissen}, P.~E. 2016, \aap, 593, A65

\bibitem[{{Pont} \& {Eyer}(2004)}]{pont04}
{Pont}, F. \& {Eyer}, L. 2004, \mnras, 351, 487

\bibitem[{{Pont} {et~al.}(2011){Pont}, {Husnoo}, {Mazeh}, \&
  {Fabrycky}}]{pont11}
{Pont}, F., {Husnoo}, N., {Mazeh}, T., \& {Fabrycky}, D. 2011, \mnras, 414,
  1278

\bibitem[{{Pont} {et~al.}(2006){Pont}, {Zucker}, \& {Queloz}}]{pont06}
{Pont}, F., {Zucker}, S., \& {Queloz}, D. 2006, \mnras, 373, 231

\bibitem[{{Prugniel} {et~al.}(2011){Prugniel}, {Vauglin}, \&
  {Koleva}}]{prugniel11}
{Prugniel}, P., {Vauglin}, I., \& {Koleva}, M. 2011, \aap, 531, A165

\bibitem[{{Ram{\'{\i}}rez} {et~al.}(2013){Ram{\'{\i}}rez}, {Allende Prieto}, \&
  {Lambert}}]{ramirez13}
{Ram{\'{\i}}rez}, I., {Allende Prieto}, C., \& {Lambert}, D.~L. 2013, \apj,
  764, 78

\bibitem[{Rasmussen \& Williams(2005)}]{rasmussen05}
Rasmussen, C.~E. \& Williams, C. K.~I. 2005, Gaussian Processes for Machine
  Learning (Adaptive Computation and Machine Learning) (The MIT Press)

\bibitem[{{Rauer} {et~al.}(2014){Rauer}, {Catala}, {Aerts}, {Appourchaux},
  {Benz}, {Brandeker}, {Christensen-Dalsgaard}, {Deleuil}, {Gizon}, {Goupil},
  {G{\"u}del}, {Janot-Pacheco}, {Mas-Hesse}, {Pagano}, {Piotto}, {Pollacco},
  {Santos}, {Smith}, {Su{\'a}rez}, {Szab{\'o}}, {Udry}, {Adibekyan}, {Alibert},
  {Almenara}, {Amaro-Seoane}, {Eiff}, {Asplund}, {Antonello}, {Barnes},
  {Baudin}, {Belkacem}, {Bergemann}, {Bihain}, {Birch}, {Bonfils}, {Boisse},
  {Bonomo}, {Borsa}, {Brand{\~a}o}, {Brocato}, {Brun}, {Burleigh}, {Burston},
  {Cabrera}, {Cassisi}, {Chaplin}, {Charpinet}, {Chiappini}, {Church},
  {Csizmadia}, {Cunha}, {Damasso}, {Davies}, {Deeg}, {D{\'{\i}}az}, {Dreizler},
  {Dreyer}, {Eggenberger}, {Ehrenreich}, {Eigm{\"u}ller}, {Erikson}, {Farmer},
  {Feltzing}, {de Oliveira Fialho}, {Figueira}, {Forveille}, {Fridlund},
  {Garc{\'{\i}}a}, {Giommi}, {Giuffrida}, {Godolt}, {Gomes da Silva},
  {Granzer}, {Grenfell}, {Grotsch-Noels}, {G{\"u}nther}, {Haswell}, {Hatzes},
  {H{\'e}brard}, {Hekker}, {Helled}, {Heng}, {Jenkins}, {Johansen},
  {Khodachenko}, {Kislyakova}, {Kley}, {Kolb}, {Krivova}, {Kupka}, {Lammer},
  {Lanza}, {Lebreton}, {Magrin}, {Marcos-Arenal}, {Marrese}, {Marques},
  {Martins}, {Mathis}, {Mathur}, {Messina}, {Miglio}, {Montalban}, {Montalto},
  {Monteiro}, {Moradi}, {Moravveji}, {Mordasini}, {Morel}, {Mortier},
  {Nascimbeni}, {Nelson}, {Nielsen}, {Noack}, {Norton}, {Ofir}, {Oshagh},
  {Ouazzani}, {P{\'a}pics}, {Parro}, {Petit}, {Plez}, {Poretti}, {Quirrenbach},
  {Ragazzoni}, {Raimondo}, {Rainer}, {Reese}, {Redmer}, {Reffert},
  {Rojas-Ayala}, {Roxburgh}, {Salmon}, {Santerne}, {Schneider}, {Schou},
  {Schuh}, {Schunker}, {Silva-Valio}, {Silvotti}, {Skillen}, {Snellen}, {Sohl},
  {Sousa}, {Sozzetti}, {Stello}, {Strassmeier}, {{\v S}vanda}, {Szab{\'o}},
  {Tkachenko}, {Valencia}, {Van Grootel}, {Vauclair}, {Ventura}, {Wagner},
  {Walton}, {Weingrill}, {Werner}, {Wheatley}, \& {Zwintz}}]{rauer14}
{Rauer}, H., {Catala}, C., {Aerts}, C., {et~al.} 2014, Experimental Astronomy,
  38, 249

\bibitem[{Roberts {et~al.}(1997)Roberts, Gelman, \& Gilks}]{roberts97}
Roberts, G.~O., Gelman, A., \& Gilks, W.~R. 1997, Ann. Appl. Probab., 7, 110

\bibitem[{{Rodr{\'\i}guez} \& {Ferraz-Mello}(2010)}]{rodriguez10}
{Rodr{\'\i}guez}, A. \& {Ferraz-Mello}, S. 2010, in EAS Publications Series,
  Vol.~42, EAS Publications Series, ed. K.~{Go{\.z}dziewski}, A.~{Niedzielski},
  \& J.~{Schneider}, 411--418

\bibitem[{{Rodr{\'\i}guez} {et~al.}(2011){Rodr{\'\i}guez}, {Ferraz-Mello},
  {Michtchenko}, {Beaug{\'e}}, \& {Miloni}}]{rodriguez11}
{Rodr{\'\i}guez}, A., {Ferraz-Mello}, S., {Michtchenko}, T.~A., {Beaug{\'e}},
  C., \& {Miloni}, O. 2011, \mnras, 415, 2349

\bibitem[{{Rosenfield} {et~al.}(2016){Rosenfield}, {Marigo}, {Girardi},
  {Dalcanton}, {Bressan}, {Williams}, \& {Dolphin}}]{rosenfield16}
{Rosenfield}, P., {Marigo}, P., {Girardi}, L., {et~al.} 2016, \apj, 822, 73

\bibitem[{{Rouan} {et~al.}(1999){Rouan}, {Baglin}, {Barge}, {Copet}, {Deleuil},
  {Leger}, {Schneider}, {Toublanc}, \& {Vuillemin}}]{rouan99}
{Rouan}, D., {Baglin}, A., {Barge}, P., {et~al.} 1999, Physics and Chemistry of
  the Earth C, 24, 567

\bibitem[{{Rybicki} \& {Lightman}(1979)}]{rybicki79}
{Rybicki}, G.~B. \& {Lightman}, A.~P. 1979, {Radiative processes in
  astrophysics}

\bibitem[{{Sanchis-Ojeda} {et~al.}(2011){Sanchis-Ojeda}, {Winn}, {Holman},
  {Carter}, {Osip}, \& {Fuentes}}]{sanchisOjeda11}
{Sanchis-Ojeda}, R., {Winn}, J.~N., {Holman}, M.~J., {et~al.} 2011, \apj, 733,
  127

\bibitem[{Schwarz(1978)}]{schwarz78}
Schwarz, G. 1978, The Annals of Statistics, 6, 461

\bibitem[{{Scuflaire} {et~al.}(2008){Scuflaire}, {Th{\'e}ado}, {Montalb{\'a}n},
  {Miglio}, {Bourge}, {Godart}, {Thoul}, \& {Noels}}]{scuflaire08}
{Scuflaire}, R., {Th{\'e}ado}, S., {Montalb{\'a}n}, J., {et~al.} 2008, \apss,
  316, 83

\bibitem[{{Seager} \& {Mall{\'e}n-Ornelas}(2003)}]{seager03}
{Seager}, S. \& {Mall{\'e}n-Ornelas}, G. 2003, \apj, 585, 1038

\bibitem[{{Sharma} {et~al.}(2018){Sharma}, {Stello}, {Buder}, {Kos},
  {Bland-Hawthorn}, {Asplund}, {Duong}, {Lin}, {Lind}, {Ness}, {Huber},
  {Zwitter}, {Traven}, {Hon}, {Kafle}, {Khanna}, {Saddon}, {Anguiano}, {Casey},
  {Freeman}, {Martell}, {De Silva}, {Simpson}, {Wittenmyer}, \&
  {Zucker}}]{sharma18}
{Sharma}, S., {Stello}, D., {Buder}, S., {et~al.} 2018, \mnras, 473, 2004

\bibitem[{{Siverd} {et~al.}(2012){Siverd}, {Beatty}, {Pepper}, {Eastman},
  {Collins}, {Bieryla}, {Latham}, {Buchhave}, {Jensen}, {Crepp}, {Street},
  {Stassun}, {Gaudi}, {Berlind}, {Calkins}, {DePoy}, {Esquerdo}, {Fulton},
  {F{\H{u}}r{\'e}sz}, {Geary}, {Gould}, {Hebb}, {Kielkopf}, {Marshall},
  {Pogge}, {Stanek}, {Stefanik}, {Szentgyorgyi}, {Trueblood}, {Trueblood},
  {Stutz}, \& {van Saders}}]{siverd12}
{Siverd}, R.~J., {Beatty}, T.~G., {Pepper}, J., {et~al.} 2012, \apj, 761, 123

\bibitem[{{Skumanich}(1972)}]{skumanich72}
{Skumanich}, A. 1972, \apj, 171, 565

\bibitem[{{Slumstrup} {et~al.}(2017){Slumstrup}, {Grundahl}, {Brogaard},
  {Thygesen}, {Nissen}, {Jessen-Hansen}, {Van Eylen}, \&
  {Pedersen}}]{slumstrup17}
{Slumstrup}, D., {Grundahl}, F., {Brogaard}, K., {et~al.} 2017, \aap, 604, L8

\bibitem[{{Smith} {et~al.}(2017){Smith}, {Gandolfi}, {Barrag{\'a}n}, {Bowler},
  {Csizmadia}, {Endl}, {Fridlund}, {Grziwa}, {Guenther}, {Hatzes}, {Nowak},
  {Albrecht}, {Alonso}, {Cabrera}, {Cochran}, {Deeg}, {Cusano},
  {Eigm{\"u}ller}, {Erikson}, {Hidalgo}, {Hirano}, {Johnson}, {Korth}, {Mann},
  {Narita}, {Nespral}, {Palle}, {P{\"a}tzold}, {Prieto-Arranz}, {Rauer},
  {Ribas}, {Tingley}, \& {Wolthoff}}]{smith17}
{Smith}, A.~M.~S., {Gandolfi}, D., {Barrag{\'a}n}, O., {et~al.} 2017, \mnras,
  464, 2708

\bibitem[{{Soderblom}(2010)}]{soderblom10}
{Soderblom}, D.~R. 2010, \araa, 48, 581

\bibitem[{{Soubiran} {et~al.}(2008){Soubiran}, {Bienayme}, {Mishenina}, \&
  {Kovtyukh}}]{soubiran08}
{Soubiran}, C., {Bienayme}, O., {Mishenina}, T.~V., \& {Kovtyukh}, V.~V. 2008,
  VizieR Online Data Catalog, 348

\bibitem[{{Southworth} {et~al.}(2009){Southworth}, {Hinse}, {Burgdorf},
  {Dominik}, {Hornstrup}, {J{\o}rgensen}, {Liebig}, {Ricci}, {Th{\"o}ne},
  {Anguita}, {Bozza}, {Novati}, {Harps{\o}e}, {Mancini}, {Masi}, {Mathiasen},
  {Rahvar}, {Scarpetta}, {Snodgrass}, {Surdej}, \& {Zub}}]{southworth09}
{Southworth}, J., {Hinse}, T.~C., {Burgdorf}, M.~J., {et~al.} 2009, \mnras,
  399, 287

\bibitem[{{Sozzetti} {et~al.}(2007){Sozzetti}, {Torres}, {Charbonneau},
  {Latham}, {Holman}, {Winn}, {Laird}, \& {O'Donovan}}]{sozzetti07}
{Sozzetti}, A., {Torres}, G., {Charbonneau}, D., {et~al.} 2007, \apj, 664, 1190

\bibitem[{{Takeda} {et~al.}(2007){Takeda}, {Ford}, {Sills}, {Rasio}, {Fischer},
  \& {Valenti}}]{takeda07}
{Takeda}, G., {Ford}, E.~B., {Sills}, A., {et~al.} 2007, \apjs, 168, 297

\bibitem[{{Ter Braak}(2006)}]{terBraak06}
{Ter Braak}, C. J.~F. 2006, Statistics and Computing, 16, 239

\bibitem[{{Torres} {et~al.}(2010){Torres}, {Andersen}, \&
  {Gim{\'e}nez}}]{torres10}
{Torres}, G., {Andersen}, J., \& {Gim{\'e}nez}, A. 2010, \aapr, 18, 67

\bibitem[{{Valenti} \& {Fischer}(2005)}]{valenti05}
{Valenti}, J.~A. \& {Fischer}, D.~A. 2005, \apjs, 159, 141

\bibitem[{{van Leeuwen}(2007)}]{vanLeeuwen07}
{van Leeuwen}, F. 2007, \aap, 474, 653

\bibitem[{Vats \& Knudson(2018)}]{vats18}
Vats, D. \& Knudson, C. 2018, arXiv e-prints [\eprint[arXiv]{1812.09384}]

\bibitem[{{Vogt} {et~al.}(2015){Vogt}, {Burt}, {Meschiari}, {Butler}, {Henry},
  {Wang}, {Holden}, {Gapp}, {Hanson}, {Arriagada}, {Keiser}, {Teske}, \&
  {Laughlin}}]{vogt15}
{Vogt}, S.~S., {Burt}, J., {Meschiari}, S., {et~al.} 2015, \apj, 814, 12

\bibitem[{{White} {et~al.}(2018){White}, {Huber}, {Mann}, {Casagrande},
  {Grunblatt}, {Justesen}, {Silva Aguirre}, {Bedding}, {Ireland}, {Schaefer},
  \& {Tuthill}}]{white18}
{White}, T.~R., {Huber}, D., {Mann}, A.~W., {et~al.} 2018, \mnras, 477, 4403

\bibitem[{{Wilson} {et~al.}(2008){Wilson}, {Gillon}, {Hellier}, {Maxted},
  {Pepe}, {Queloz}, {Anderson}, {Collier Cameron}, {Smalley}, {Lister},
  {Bentley}, {Blecha}, {Christian}, {Enoch}, {Haswell}, {Hebb}, {Horne},
  {Irwin}, {Joshi}, {Kane}, {Marmier}, {Mayor}, {Parley}, {Pollacco}, {Pont},
  {Ryans}, {Segransan}, {Skillen}, {Street}, {Udry}, {West}, \&
  {Wheatley}}]{wilson08}
{Wilson}, D.~M., {Gillon}, M., {Hellier}, C., {et~al.} 2008, \apjl, 675, L113

\bibitem[{{Winn}(2010)}]{winn10}
{Winn}, J.~N. 2010, arXiv e-prints [\eprint[arXiv]{1001.2010}]

\bibitem[{{Winn} {et~al.}(2009){Winn}, {Holman}, {Carter}, {Torres}, {Osip}, \&
  {Beatty}}]{winn09}
{Winn}, J.~N., {Holman}, M.~J., {Carter}, J.~A., {et~al.} 2009, \aj, 137, 3826

\bibitem[{{Wright}(2018)}]{wright18}
{Wright}, J.~T. 2018, {Radial Velocities as an Exoplanet Discovery Method}, 4

\bibitem[{{Wright} {et~al.}(2004){Wright}, {Marcy}, {Butler}, \&
  {Vogt}}]{wright04}
{Wright}, J.~T., {Marcy}, G.~W., {Butler}, R.~P., \& {Vogt}, S.~S. 2004, \apjs,
  152, 261

\bibitem[{{Yi} {et~al.}(2001){Yi}, {Demarque}, {Kim}, {Lee}, {Ree}, {Lejeune},
  \& {Barnes}}]{yi01}
{Yi}, S., {Demarque}, P., {Kim}, Y.-C., {et~al.} 2001, \apjs, 136, 417

\end{thebibliography}
\bibliographystyle{aa}

\begin{appendix}
\section{Relation between transit duration $W$ and the scaled semi-major axis $\frac{a}{R_{\star}}$}
 For the transit analysis our code computes the Cartesian coordinates of the Keplerian orbit and then it feeds the \citet{mandel02} model with the planet to star sky-projected distance and the $\frac{R_p}{R_{\star}}$ size ratio. Those Cartesian coordinates are computed using notably $\frac{a}{R_{\star}}$; relying on transit observables, $\frac{a}{R_{\star}}$ comes from $P$, $W$, $b$, and $\mathrm{d}F$ (besides $e$ and $\omega$ if the orbit is eccentric).
  
 Following \citet{winn10}, establishing the total transit duration $W$, that is the temporal interval between the first and fourth contact requires us to compute the integral
 \begin{equation}
  W=\frac{P}{2\pi\sqrt{1-e^2}}\int_{f_{\mathrm{I}}}^{f_{\mathrm{IV}}}\frac{r^2(f)}{a^2}\mathrm{d}f
  \label{eq:W}
 ,\end{equation}
 where $r(f)$ is the ellipse equation (i.e. the star-planet distance) as a function of the true anomaly $f$
 \begin{equation}
  r=\frac{a(1-e^2)}{1+e\cos{f}}
  \label{eq:r}
 ,\end{equation}
 while $f_{\mathrm{I}}$ and $f_{\mathrm{IV}}$ are the true anomalies at the moment of first and last contact, respectively. By projecting $r$ onto the plane of the sky from eq. (53)-(54) by \citet{murray11}, we obtain the star-planet distance in the sky plane, that is
 \begin{equation}
  r_{\mathrm{sky}}=\frac{a(1-e^2)}{1+e\cos{f}}\sqrt{1-\sin^2{(\omega+f)}\sin{i_p}}
  \label{eq:rSky}
 .\end{equation}
 By imposing $r_{\mathrm{sky}}=R_{\star}+R_p$ and then solving eq. (\ref{eq:rSky}) in terms of $f$, $f_{\mathrm{I}}$, and $f_{\mathrm{IV}}$ are recovered, such that $W$ can be finally computed from (\ref{eq:W}).
 
 It is clear that all this procedure leads to lengthy algebra and to numerical resolution of equations. Nonetheless, there are very useful approximations which relate $W$ to transit observables in a more straightforward way, and the $W$-$\frac{a}{R_{\star}}$ relation that has been implemented in our algorithm is written as
 \begin{equation}
  \frac{a}{R_{\star}}=\frac{P}{\pi W}\tilde{e}_1\sqrt{(1+\sqrt{\mathrm{d}F})^2 - (b'\tilde{e}_2)^2}
  \label{eq:aR}
 ,\end{equation}
 where
 \begin{equation}
  \tilde{e}_1=\frac{\sqrt{1-e^2}}{1+e\sin{\omega}}
  \label{eq:e1}
 \end{equation}
 \begin{equation}
  \tilde{e}_2=\frac{1-e^2}{1+e\sin{\omega}}
  \label{eq:e2}
 .\end{equation}
 Eq. (\ref{eq:aR}) is a revised version of eq. (8) by \citet{seager03}, where we added the eccentricity-dependent factors (\ref{eq:e1}) and (\ref{eq:e2}), as suggested by \citet{winn10}, and we considered the following approximations:
\begin{equation}
\sin{\frac{\pi W}{P}}\approx\frac{\pi W}{P}\quad\quad\quad \cos^2{\frac{\pi W}{P}}\approx1.
\label{eq:approxSinCos}
\end{equation}

We evaluated the impact of our approximations, also thanks to some reference data of interest that we inferred from the NASA Exoplanet archive. It turned out that $\frac{a}{R_{\star}}$ values which are computed through (\ref{eq:aR}) differ from those derived from the actual formula of \citet{seager03} by less than $1\%$ if $\frac{W}{P}<0.07$, which holds for more than $97\%$ of the exoplanets that have been confirmed so far. 
It is even more reassuring that the median value of $\frac{W}{P}=0.015$ produces negligible variations on $\frac{a}{R_{\star}}$ ($\sim0.05\%$), while in the rare cases in which $\frac{W}{P}>0.07$, we may register variations on $\frac{a}{R_{\star}}$ of order of a few percent, which are still not worrying in general since the median relative uncertainty on $\frac{a}{R_{\star}}$ is $\sim10\%$.

Without paying a high price, the introduced approximations enable us to easily express $W$ as a function of $\frac{a}{R_{\star}}$. This is useful in case the transit is very shallow and it is preferable not to rely on it to infer $W$. As also explained in \S\ref{ssec:MCMCI}, in this scenario $\rho_{\star}$ is established from evolutionary models, then $\frac{a}{R_{\star}}$ is computed from $\rho_{\star}$ thanks to Kepler's third law, and finally also $W$ is available by easily inverting eq. (\ref{eq:aR}).
  
\end{appendix}


\end{document}